\documentclass[authoryear,review,12pt,authoryear]{elsarticle}

\usepackage{amssymb}
\usepackage[mathscr]{eucal}
\usepackage{amsmath}
\usepackage{amsfonts,bbm}
\usepackage{mathrsfs} 
\usepackage{color}
\usepackage{ragged2e}
\usepackage{tikz} 
\usepgflibrary{arrows}
\usetikzlibrary{calc}
\usetikzlibrary{patterns}
\newdefinition{remark}{Remark}
\begin{document}
\begin{frontmatter}
\title{Equilibrium of Kirchhoff's rods subject to a distribution of magnetic couples}  
\author{Marzio Lembo}
\ead{marzio.lembo@uniroma3.it}
\author{Giuseppe Tomassetti} 
\ead{giuseppe.tomassetti@uniroma3.it}
\affiliation{organization={Dipartimento di Ingegneria, Universit\`a di Roma Tre},
            addressline={Via Vito Volterra 62}, 
            city={Roma},
            postcode={00146}, 
            country={Italy}} 
\begin{abstract}
The equilibrium of magneto-elastic rods, formed of an elastic matrix containing a uniform distribution of paramagnetic particles, that are subject to terminal loads and are immersed in a uniform magnetic field, is studied. 
The deduced nonlinear equilibrium equations are fully consistent with Kirchhoff's theory in the sense that they hold at the same order of magnitude. 
Exact solutions of those equations in terms of Weierstrass elliptic functions are presented with reference to magneto-elastic cantilevers that undergo planar deformations under the action of a terminal force and a magnetic field whose directions are either parallel or orthogonal. 
The exact solutions are applied to the study of a problem of remotely controlled deformation of a rod and to a bifurcation problem in which the end force and the magnetic field act as an imperfection parameter and a bifurcation parameter, respectively. 
\end{abstract} 
%
%
%
%
\end{frontmatter}
\section{Introduction} 
The theory of Kirchhoff describes nonlinear deformations of thin rods in which the displacements may be large while the strains with respect to an undistorted configuration remain small. 
Starting from a three-dimensional model of a rod, Kirchhoff's theory arrives at one-dimensional equations of motion that are exact at the first order in a dimensionless parameter depending on the thickness of the rod, the curvature of the rod axis and the twist in both the undistorted and the deformed configurations, and the extension suffered by the rod axis.

In the present paper, we examine the equilibrium of elastic rods containing a uniform distribution of paramagnetic particles, that are subject to terminal loads and are immersed in a uniform magnetic field. 
In a (three-dimensional) body of such a type, the magnetic field determines the magnetization of the embedded particles, and the interaction of the magnetized particles with the applied magnetic field produces a distribution of magnetic couples acting on the body. 
For that problem, we reduce the magnetic action to a distribution of couples per unit length of the rod axis and deduce the equilibrium equations with a procedure that is completely coherent with the theory of Kirchhoff because the obtained results hold at the same order of magnitude of the theory.  
Then we apply the equilibrium equations to the study of planar deformations of cantilevers immersed in a magnetic field and acted upon by a force applied at their free end, and obtain exact solutions in terms of Weierstrass elliptic functions for the cases in which the terminal force and the magnetic field are either parallel or orthogonal. 
Two examples of application of these exact solutions are presented, in which we have focused our attention on the dependence of the deformation on the magnetic field when the force is kept constant. 
The first example, in which the field and the force have orthogonal directions, may be viewed as a model of a device for the remote control of the deformation of a rod; the second one, in which the directions of the force and the field are parallel, shows the presence of bifurcation points and produces bifurcation diagrams in which the force has the role of an imperfection parameter while the magnetic field act as a bifurcation parameter. 
The examples suggest that magneto-elastic rods may exhibit a wide variety of behaviors when the parameters on which the deformation depends are changed. 

The theory of bending and twisting of thin rods has been developed by \citet{KG1859, KG1876} and \citet{CA1862, CA1883}, and refined by \cite{AL1944}, who justified some passages of the theory by means of an order of magnitude analysis and showed that it is a first-order theory in an appropriate measure of thickness, curvature, and extension. 
The development of the theory and the contributions to it of various mathematicians of the nineteenth century, are presented in the Introduction of the cited Love's treatise and in a historical paper of \cite{ED1992}; a concise presentation of the theory in a modern notation has been given by \cite{BC1993}.  
The interest for Kirchhoff's theory in various fields of mechanics may be ascribed also to the possibility of finding exact solutions of its nonlinear equations in terms of elliptic functions. 
Among the first applications of the theory are those to the elastica and the stability of rods, based on the Kirchhoff's theorem of the kinetic analogue.    
Recently, the theory has been applied in molecular biology to the study of equilibrium, motions, and stability of segments of DNA \citep[e.g.,][]{TCO1994, TCL1996, CS2000} and in mechanics of nonlocal elastic material to the study of equilibrium, stability, and buckling of nanorods \citep{ML2016, ML2017, ML2018}. 

The mechanics of rods under magnetic actions is the subject of a growing field of research. 
Without attempting an account of the literature on the matter, we restrict ourself to recall that the coupling between elastic and magnetic phenomena in bodies modelled as wires, beams, or rods has been initially studied with reference to the stability of beams made of magnetic materials and conducting wires and rods immersed in a magnetic field \citep[e.g.][]{MP1969, WP1972, MH1979, PW1983, SW1988, VP2013}. 
In the last years, developments in material science have led to a great interest in composite bodies whose shape can be remotely controlled by means of a magnetic field \citep[e.g.][]{GF2003, DB2005, CC2007, KU2012, GW2015}. 
In particular, \citet{CF2017} have derived a model for fiber-reinforced magneto-elastic bodies, containing a uniform distribution of prolate paramagnetic particles firmly embedded in an elastomeric matrix, in which the action of a uniform magnetic field induces distributions of magnetic couples.   
The expression obtained in this model for the density of magnetostatic energy in a three-dimensional body is the starting point for the present deduction of a distribution of couples per unit length of the rod axis that is consistent, in the sense above specified, with Kirchhoff's theory of rods.  
Among recent studies on deformations of rods controlled by a magnetic field, we recall the paper of \citet{WK2020}, in which large deformations of hard-magnetic elastica subject to magnetic forces and couples have been discussed with a view toward application to the design and control of small-scale robots employed in diagnostic and therapeutic medical procedures, and the paper of \citet{DGT2020}, in which a problem of optimal design and control of shape for a non-uniformly magnetized cantilever subject to an external magnetic field has been formulated and solved.

The present paper is organized as follows.  
In the Section 2, after giving a summary of some results of Kirchhoff's theory, the equilibrium equations for magneto-elastic rods of the type previously described are deduced. 
Section 3 is devoted to the study of the equilibrium equations of magneto-elastic cantilevers that undergo planar deformations under the action of a terminal force and a uniform magnetic field; exact solutions, in terms of Weierstrass elliptic functions, are obtained for the cases in which the field and the force have parallel or orthogonal directions. 
Section 4 presents the two above mentioned examples of application of these exact solutions, and Section 5 contains some concluding remarks.     
In the present paper, the convention of the sum on repeated indices is adopted, with the agreement that latin indices range over $\{1, 2, 3\}$, and greek indices over $\{1, 2\}$.

\section{Equilibrium of Kirchhoff's rods subject to magnetic couples} \label{S:eKr}

To derive the equilibrium equations of magneto-elastic rods which deform according to Kirchhoff's theory, we shall make use of an expression of the density of magnetostatic energy that depends on the relative orientation of the particles and the magnetic field, with the orientation of the particles determined by the gradient of the deformation.   
In the present Section, after having recalled some features of Kirchhoff's theory that are of interest for the following discussion, we determine the form that the density of magnetostatic energy assumes when the deformation of a rod is described by Kirchhoff's theory.  
Then we deduce the equilibrium equations of a magneto-elastic rod by taking the variation of its elastic and magnetic energy.

\subsection{Kirchhoff's theory of rods} \label{S:Ktr}

We begin by recalling those results of Kirchhoff's theory that will be employed in the following analysis of magneto-elastic rods. 
We refer the reader interested in a more detailed or general treatment of the subject to the original works and the modern expositions cited in the Introduction.  

In Kirchhoff's theory, a rod $\mathscr{R}$ is seen as a three-dimensional body that in an undistorted stress-free configuration $\mathfrak{C}^\mathrm{u}$ occupies a space region described by the mapping 
\begin{equation} \label{kthr1} 
\hat{\boldsymbol x}^{\mathrm{u}}\left(X_{1}, X_{2},s\right)={\boldsymbol x}^{\mathrm{u}}(s)+X_{1} \boldsymbol d_{1}^{\mathrm{u}}(s)+X_{2} \boldsymbol d_{2}^{\mathrm{u}}(s),
\end{equation}
where: 
$\boldsymbol x^{\mathrm{u}}(s)$ is the position on a smooth space curve $\mathscr{C}^\mathrm{u}$ of the point having arc-length coordinate $s$, with $s$ varying in $(0,\ell)$; 
$X_1$ and $X_2$ belong to a connected domain $\Sigma$ of $\mathbb R^2$ whose centroid coincides with origin; ${\boldsymbol d}^{\rm u}_{1}(s)$ and ${\boldsymbol d}^{\rm u}_{2}(s)$ are smooth unit vectors orthogonal to $\mathscr{C}^\mathrm{u}$ at $\boldsymbol x^{\mathrm{u}}(s)$, that, together with the unit tangent ${\boldsymbol d}^{\rm u}_{3}(s)$ to $\mathscr{C}^\mathrm{u}$, 
\begin{equation} \label{kthr2} \nonumber 
\boldsymbol d^{\mathrm u}_3(s)={\boldsymbol x^{\mathrm u}}'(s)=\frac {d \boldsymbol x^{\mathrm u}}{{d}s} (s) \,,
\end{equation}
form a right-handed orthonormal triad.   
The set of points $\hat{\boldsymbol x}^{\mathrm{u}}\left(X_{1}, X_{2},s\right)$ with $s$ fixed is the \emph{cross section} $\mathscr{S} (s)$ of $\mathscr{R}$; the unit vectors ${\boldsymbol d}^{\rm u}_{1}(s)$ and ${\boldsymbol d}^{\rm u}_{2}(s)$ are chosen lying along the principal directions of inertia of $\mathscr{S} (s)$.  
The curve $\mathscr{C}^\mathrm{u}$ is the locus of the centroids of the cross sections, and the set of particles of $\mathscr{R}$ that are on $\mathscr{C}^\mathrm{u}$ form the \emph{axis} of the rod.  
The number $\ell$ is the \emph{length} of $\mathscr{R}$; denoted by $h$ the maximum distance between the origin and points on the boundary of $\Sigma$, the body $\mathscr{R}$ is called a ``rod'' provided that $h/\ell \ll 1$.  

As it is usual in applications of Kirchhoff's theory, we assume that the rod is \emph{inextensible}, that is, in its deformations we neglect the extensions of the axis; such assumption is justified because those extensions ordinarily are small with respect to the other quantities that determine  the order of magnitude of the approximations of the theory \citep[cf.][Sect. 258]{AL1944}.  

In a deformation of the rod from $\mathfrak{C}^{\mathrm{u}}$ to a configuration $\mathfrak{C}$, the curve $\mathscr{C}^\mathrm{u}$ is transformed into a curve $\mathscr{C}$, given by an equation of the form $\boldsymbol{x} = \boldsymbol{x} (s)$. 
In view of the assumption of inextensibility, $s$ is an arc-length parameter on $\mathscr{C}$ and the vector $\boldsymbol{d}_3 (s) = \boldsymbol{x}^\prime (s)$ is the unit tangent to $\mathscr{C}$ at $\boldsymbol{x} (s)$; thus, in each deformation, the following \emph{inextensibility condition} holds: 
\begin{equation} \label{kthr3}
\boldsymbol{d}_3 (s) \cdot \boldsymbol{d}_3 (s) = \boldsymbol{x}^\prime (s) \cdot \boldsymbol{x}^\prime (s) = 1 \,. 
\end{equation}
The vectors $\boldsymbol{d}^\mathrm{u}_{1}(s)$ and $\boldsymbol{d}^{\rm u}_{2}(s)$ are transformed into the vectors $\boldsymbol{d}_1 (s)$ and $\boldsymbol{d}_2 (s)$ that are tangent, at $\boldsymbol{x} (s)$, to the curves in which the principal axes of inertia of $\mathscr{S} (s)$ have been transformed. 
To within the approximation of the theory,  $\boldsymbol{d}_1 (s)$ and $\boldsymbol{d}_2 (s)$ can be regarded as orthogonal, of unit length, and lying in the plane perpendicular to $\boldsymbol{d}_3 (s)$ \citep[cf.][Sect.4]{ED1992}.  
Since the triad $(\boldsymbol{d}_1, \boldsymbol{d}_2, \boldsymbol{d}_3)$ is orthonormal, in each configuration of the rod there exists a vector $\boldsymbol{\kappa} = \boldsymbol{\kappa}(s) = \kappa_i (s) \boldsymbol{d}_i (s)$, called the \emph{curvature vector}, such that 
\begin{equation} \label{kthr4} 
\boldsymbol{d}_i^\prime (s) = \boldsymbol{\kappa} (s) \times \boldsymbol{d}_i (s) \,, \quad i = 1, 2, 3 \,; \quad  \qquad  \boldsymbol{\kappa} (s) = \frac{1}{2} \boldsymbol{d}_i (s) \times \boldsymbol{d}_i^\prime (s) \,. 
\end{equation}
The functions $\kappa_1 = \kappa_1 (s)$ and $\kappa_2 = \kappa_2 (s)$ are the \emph{components of curvature}; the relationship between these quantities and the geometric curvature $k = k(s)$ of $\mathscr{C}$ is made clear by the equations  
\begin{equation} \label{kthr5}  
\kappa_1 \boldsymbol{d}_1  + \kappa_2 \boldsymbol{d}_2 = k \boldsymbol{b} = \boldsymbol{d}_3 \times \boldsymbol{d}_3^{\, \prime} \,, \qquad \quad k = \sqrt{ \kappa_1^2 + \kappa_2^2  } \,, 
\end{equation}
where $\boldsymbol{b} = \boldsymbol{b}(s)$ is the binormal of the curve $\mathscr{C}$.  
The function $\kappa_3 = \kappa_3 (s)$ is the \emph{twist density} in the configuration $\mathfrak{C}$, and is related to the geometric torsion $\tau = \tau (s)$ of $\mathscr{C}$ through  
\begin{equation} \label{kthr6} \nonumber 
\kappa_3 = \tau + \phi^\prime  \,,
\end{equation}
where $\phi = \phi(s)$ is the angle between the principal normal $\boldsymbol{n}$ of $\mathscr{C}$ and $\boldsymbol{d}_1$. 

Now it is possible to make precise the order of magnitude of the approximations of the theory. 
Let $\boldsymbol{\kappa}^\mathrm{u} = \kappa_i^\mathrm{u} \boldsymbol{d}_i^\mathrm{u}$ denote the curvature vector for the configuration $\mathfrak{C}^\mathrm{u}$, and let $\varepsilon$ be defined by  
\begin{equation} \label{kthr7} 
\varepsilon = \max_{s\in(0,\ell)}\{ | \boldsymbol{\kappa}^\mathrm{u}(s) | h, \, | \boldsymbol{\kappa}(s) | h, \, h/\ell\} \,;
\end{equation}
the theory of Kirchhoff holds to within an error of order $O(\varepsilon^2)$.   

According to Love's approach to the theory \citep[][Sect.256]{AL1944}, a deformation of the rod from the undistorted configuration $\mathfrak{C}^\mathrm{u}$ to a configuration $\mathfrak{C}$ can be imagined as obtained by taking the rod in a state in which the cross sections remain plane, unstrained, and orthogonal to the axial curve $\mathscr{C}$, and suffer a rotation corresponding to the right value of the twist, and by adding to that state a ``small'' displacement $\bar{\boldsymbol{u}}$,  which vanishes on $\mathscr{C}$.  
Thus, the positions $\hat{\boldsymbol{x}}$ of the points of $\mathscr{R}$ in the configuration $\mathfrak{C}$ are given by an equation of the form   
\begin{equation} \label{kthr8} 
\hat{\boldsymbol{x}} \left(X_{1}, X_{2},s\right) = {\boldsymbol x} (s)+X_{1} \boldsymbol d_{1} (s)+ X_{2} \boldsymbol d_{2} (s) + \bar{\boldsymbol{u}} \left(X_{1}, X_{2},s\right) , 
\end{equation}
where $\bar{\boldsymbol{u}} (0,0,s) = \boldsymbol{0}$, and 
\begin{equation} \label{kthr9} 
\frac{| \bar{\boldsymbol{u}} |}{h} = O (\varepsilon) \,, \qquad \left| \frac{ \partial \bar{\boldsymbol{u}} }{\partial X_\alpha} \right| = O (\varepsilon)\,, \qquad \left| \frac{\partial \bar{\boldsymbol{u}}} {\partial s} \right| = O (\varepsilon^2) \,. 
\end{equation}
The maps $\hat{\boldsymbol{x}}^\mathrm{u}$ and $\hat{\boldsymbol{x}}$, given by equations \eqref{kthr1} and \eqref{kthr8}, are defined for $(X_1, X_2 , s)$ in the cylinder $\mathcal{C} = \Sigma \times (0, \ell)$ of $\mathbbm{R}^3$. 
The triplets $(X_1, X_2 , s)$ can be regarded as convected coordinates on $\mathscr{R}$; then, differentiation of equation \eqref{kthr1} yields that the vectors $(\boldsymbol{g}_1, \boldsymbol{g}_2, \boldsymbol{g}_3)$ of the covariant basis in the configuration $\mathfrak{C}^\mathrm{u}$ are: 
\begin{equation} \label{kthr10} 
\boldsymbol{g}_1 = \frac{\partial \hat{\boldsymbol{x}}^\mathrm{u}}{\partial X_1} = \boldsymbol{d}^\mathrm{u}_1 \,, \quad 
\boldsymbol{g}_2 = \frac{\partial \hat{\boldsymbol{x}}^\mathrm{u}}{\partial X_2} = \boldsymbol{d}^\mathrm{u}_2 \,, \quad 
\boldsymbol{g}_3 = \frac{\partial \hat{\boldsymbol{x}}^\mathrm{u}}{\partial s} = \boldsymbol{d}^\mathrm{u}_3 + \boldsymbol{\kappa}^\mathrm{u} \times X_\alpha \boldsymbol{d}^\mathrm{u}_\alpha \,;   
\end{equation}
the vectors $(\boldsymbol{g}^1, \boldsymbol{g}^2, \boldsymbol{g}^3)$ of the corresponding contravariant basis are 
\begin{equation} \label{kthr11} 
\boldsymbol{g}^k = \boldsymbol{d}^\mathrm{u}_k - \boldsymbol{d}^\mathrm{u}_3 \otimes \boldsymbol{d}^\mathrm{u}_k  ( X_\alpha \boldsymbol{\kappa}^\mathrm{u} \times \boldsymbol{d}^\mathrm{u}_\alpha ) + O (\varepsilon^2) \,, \qquad k = 1, 2, 3 \,. 
\end{equation}
By means of \eqref{kthr7}, \eqref{kthr8}, \eqref{kthr9}, and \eqref{kthr11}, the deformation gradient 
\begin{equation} \label{kthr12}  \nonumber 
\boldsymbol{\mathsf{F}} = \frac{\partial \hat{\boldsymbol{x}}}{\partial X_\alpha} \otimes \boldsymbol{g}^\alpha + \frac{\partial \hat{\boldsymbol{x}}}{\partial s} \otimes \boldsymbol{g}^3 \,, 
\end{equation}
can be expressed in the form 
\begin{equation} \label{kthr13}
\boldsymbol{\mathsf{F}} = \boldsymbol{Q} + X_\alpha \boldsymbol{\mu} \times \boldsymbol{d}_\alpha \otimes \boldsymbol{d}^\mathrm{u}_3 + \frac{\partial \bar{\boldsymbol{u}}}{\partial X_\alpha} \otimes \boldsymbol{d}^\mathrm{u}_\alpha + O(\varepsilon^2) \,,
\end{equation}
with    
\begin{equation} \label{kth14}
\boldsymbol{Q} =\boldsymbol d_i\otimes\boldsymbol d^{\mathrm u}_i \,,
\end{equation}
the rotation that transforms the triad $(\boldsymbol{d}^{\mathrm u}_1, \boldsymbol{d}^{\mathrm u}_2, \boldsymbol{d}^{\mathrm u}_3)$ into the triad $(\boldsymbol{d}_1, \boldsymbol{d}_2, \boldsymbol{d}_3)$, and where the vector $\boldsymbol{\mu}$, defined as   
\begin{equation} \label{kth15} 
\boldsymbol{\mu} =  \boldsymbol{\kappa} - \boldsymbol{Q} \boldsymbol{\kappa}^\mathrm{u} \,,   
\end{equation}
can be interpreted as a measure of the deformation (see equation \eqref{kth17}).  
Integration over the sections of the three-dimensional equilibrium equations written in terms of the first Piola-Kirchhoff stress tensor, yields the equilibrium equations of a rod 
\begin{equation} \label{kth16} 
\boldsymbol{F}^\prime + \boldsymbol{f} = \boldsymbol{0} \,,
\qquad \quad 
\boldsymbol{M}^\prime + \boldsymbol{d}_3 \times \boldsymbol{F} + \boldsymbol{m} = \boldsymbol{0} \,,
\end{equation}
where, at each $s$, the vectors $\boldsymbol{F} = \boldsymbol{F}(s)$ and $\boldsymbol{M} = \boldsymbol{M}(s)$ are the \emph{resultant force} and the \emph{resultant moment} with respect to $\boldsymbol{x} (s)$ of the Piola stresses exerted, on the surface in which $\mathscr{S}(s)$ has been transformed, by the part of rod on the side of increasing $s$, and $\boldsymbol{f} = \boldsymbol{f}(s)$ and $\boldsymbol{m} = \boldsymbol{m}(s)$ are the force and the couple per unit length of $\mathscr{C}^\mathrm{u}$, obtained by integration over the sections of the forces applied to the rod.  
The resultant force $\boldsymbol{F}$ is a reactive variable, not constitutively determined; to within the order of approximation of the theory, the constitutive equation for the resultant moment can be given the form  
\begin{equation} \label{kth17} 
\boldsymbol{M} = \boldsymbol{C} \boldsymbol{\mu} \,,  
\end{equation}
in which, for a rod whose cross sections have equal the two principal moments of inertia, the second-order tensor $\boldsymbol{C}$ is 
\begin{equation} \label{kth18} 
\boldsymbol{C} = EI \boldsymbol{d}_\alpha \otimes \boldsymbol{d}_\alpha + GJ \boldsymbol{d}_3 \otimes \boldsymbol{d}_3 \,, 
\end{equation}
where $E$ and $G$ are the tensile and shear moduli of the material forming the rod, $I$ is the value of the two principal moments of inertia of the cross sections, and $J$ is the torsional rigidity factor, determined by the geometry of the cross sections.   
The strain energy density $\psi_e$ per unit length of $\mathscr{C}^\mathrm{u}$ is  \citep[cf.][]{ML2003}
\begin{equation} \label{kth19}  \nonumber 
\psi_e = \psi_e (\boldsymbol{\mu}) = \frac{1}{2} \boldsymbol{M} \cdot \boldsymbol{\mu} = \frac{1}{2} \boldsymbol{C} \boldsymbol{\mu} \cdot \boldsymbol{\mu} \,;  
\end{equation}
the symmetry of $\boldsymbol{C}$ and the form of $\psi_e$ show that $d \psi_e / d \boldsymbol{\mu} = \boldsymbol{M}$. 

\subsection{Magnetostatic energy in Kirchhoff's rods} 

In the present Section we determine the expression of the magnetostatic energy in a Kirchhoff's rod formed by an elastic material containing a uniform distribution of paramagnetic particles, under the assumptions that: 

\smallskip 
\noindent 
\textit{i}) the magnetic particles have the shape of prolate ellipsoids of revolution;

\smallskip 
\noindent 
\textit{ii})  
the orientation of the particles in the undistorted configuration is specified by the unit vector $\boldsymbol a^{\mathrm u}$ giving the direction their major axis;

\smallskip 
\noindent 
\textit{iii})  the density $\nu$ of magnetic particles in the rod is uniform;

\smallskip 
\noindent 
\textit{iv})  the applied magnetic field $\mathbbm{h}_\mathrm{a}$, to which the rod is subjected, is spatially constant.

It is shown in \citet{CF2017} that, under the assumptions \textit{i})-\textit{iv}), when a magneto-elastic body undergoes a finite deformation, the magnetostatic energy $\hat{\psi}_m$ per unit volume of the reference configuration is 
\begin{equation} \label{kth20b}  
\hat{\psi}_m = - \frac{1}{2} \mu_0\nu V_\Pi \big( \chi (\mathbbm{h}_\mathrm{a} \cdot \boldsymbol{a} )^2 - \tilde\chi|\mathbbm h_{\rm a}|^2 \big) \,,    
\end{equation}
where $\mu_0$ is permeability of vacuum, $V_\Pi$ is the volume of a magnetic particle, and $\boldsymbol{a}$ is the unit vector that gives the orientation of the particles in the deformed configuration and is defined by $\boldsymbol{a} = \boldsymbol{\mathsf{F}} \boldsymbol{a}^{\mathrm u} / |\boldsymbol{\mathsf{F}} \boldsymbol{a}^{\mathrm u}|$, with $\boldsymbol{\mathsf{F}}$ the gradient of the deformation experienced by the body.  
The susceptibilities $\chi$ and $\tilde{\chi}$ are used to relate the density of magnetization $\mathbbm{m}$ to the applied field $\mathbbm{h}_\mathrm{a}$ through the equation 
\begin{equation} \label{elmr11}  
\mathbbm{m} = \chi ( \mathbbm{h}_\mathrm{a} \cdot \boldsymbol{a} ) \boldsymbol{a} + \tilde{\chi} \mathbbm{h}_\mathrm{a} \,,    
\end{equation}
and have the expressions that can be determined as follows. 

The magnetic field $\mathbbm{h}$ in a particle is the sum of the applied field $\mathbbm{h}_\mathrm{a}$ and the demagnetizing field $\mathbbm{h}_\mathrm{d}$, which is due to the magnetization within the particle and depends on its geometry: 
\begin{equation}\label{campi}
\mathbbm{h} = \mathbbm{h}_{\rm a}+\mathbbm h_{\rm d} \,. 
\end{equation} 
For a linearly magnetic material, the density of magnetization is determined by the field $ \mathbbm h$ through the equation 
\begin{equation}\label{eqazm}
\mathbbm{m} = \boldsymbol{\mathcal X} \mathbbm{h}  \,,   
\end{equation} 
where $\boldsymbol{\mathcal X}$ is the \textit{susceptibility tensor}. 
We assume that this tensor has the form 
\begin{equation}\label{suscet}
\boldsymbol{\mathcal X} = \chi_\parallel \boldsymbol{a} \otimes \boldsymbol{a} + \chi_\perp (\boldsymbol{I} - \boldsymbol{a} \otimes \boldsymbol{a}) \,,  
\end{equation} 
that describes an anisotropic material whose susceptibility is the same in all the directions orthogonal to $\boldsymbol{a}$, and has a different value in the direction $\boldsymbol{a}$;  
the constants $\chi_\parallel >0$ and $\chi_\perp>0$ are the magnetic susceptibilities in the directions parallel and orthogonal to $\boldsymbol{a}$. 
The magnetization $\mathbbm{m}$ determines the demagnetizing field $\mathbbm{h}_{\rm d}$ by means of the equation 
\begin{equation}\label{magnetiz} 
\mathbbm{h}_{\rm d} = - \boldsymbol{\mathcal N} \mathbbm{m} \,,  
\end{equation} 
where $\boldsymbol{\mathcal N}$ is the positive-definite \textit{demagnetizing tensor}, that for a prolate ellipsoid of revolution has the expression \citep[cf., e.g.,][]{HS1998} 
\begin{equation}\label{demten}
\boldsymbol{\mathcal N} = N_{\parallel} \boldsymbol{a} \otimes \boldsymbol{a} + N_\perp (\boldsymbol I -\boldsymbol a \otimes \boldsymbol a) \,,   
\end{equation}
with $N_{\parallel}$ and $N_\perp$ the demagnetizing factors along the directions parallel and orthogonal to $\boldsymbol{a}$. 
It follows from equations \eqref{campi},  \eqref{eqazm}, and \eqref{magnetiz}  that 
\begin{equation}\label{magnha} \nonumber 
\mathbbm{m} = \left( \boldsymbol{\mathcal X}^{-1} + \boldsymbol{\mathcal N} \right)^{-1}  \mathbbm{h}_{\rm a}\,; 
\end{equation}
in view of definitions \eqref{suscet} and \eqref{demten}, this equation can be written in the form \eqref{elmr11} by putting  $\chi = \big(\chi_{\|}^{-1} + N_{\|}\big)^{-1} - \big(\chi_{\perp}^{-1} + N_{\perp} \big)^{-1}$, and $\tilde{\chi} = \big( \chi_{\perp}^{-1} + N_{\perp} \big)^{-1}$. 

Our next task is to determine the dependence of the vector $\boldsymbol{a}$, and hence of the function $\hat{\psi}_m $, on the fields that describe the deformation of a Kirchhoff's rod. 
We assume that $\boldsymbol a^{\mathrm u} = \boldsymbol a^{\mathrm u} (\hat{\boldsymbol x}^{\mathrm{u}})$ is parallel to the longitudinal fibers of the rod, that is, to the material curves in $\mathfrak{C}^\mathrm{u}$ on which $X_1$ and $X_2$ do not vary, so that $\boldsymbol a^{\mathrm u} = \boldsymbol{g}_3 / | \boldsymbol{g}_3 |$. 
By making use of the expressions \eqref{kthr10}$_3$ and \eqref{kthr13} of $\boldsymbol{g}_3$ and $\boldsymbol{\mathsf{F}}$, and taking account of \eqref{kthr9}, we have that 
\begin{equation} \label{kth21}  \nonumber 
|\boldsymbol{\mathsf{F}} \boldsymbol a^{\mathrm u}|^2 = \frac{1 + 2 X_\alpha \boldsymbol{d}_3 \cdot \boldsymbol{\kappa} \times \boldsymbol{d}_\alpha + O(\varepsilon^2)}{| \boldsymbol{g}_3 |^2} \,,
\end{equation}
and
\begin{equation} \label{kth22}  \nonumber 
(\mathbbm{h}_\mathrm{a} \cdot \boldsymbol{\mathsf{F}} \boldsymbol{a}^{\mathrm u})^2 = \frac{(\mathbbm{h}_\mathrm{a} \cdot\boldsymbol d_3)^2 + 2 X_\alpha (\mathbbm{h}_\mathrm{a} \cdot \boldsymbol{\kappa} \times \boldsymbol{d}_\alpha)  (\mathbbm{h}_\mathrm{a} \cdot\boldsymbol d_3)+O(\varepsilon^2)} {| \boldsymbol{g}_3 |^2} \,.
\end{equation}
It follows from the last two equations that 
\begin{equation}  \label{kth23}  \nonumber 
\begin{split}
(\mathbbm{h}_\mathrm{a} \cdot \boldsymbol{a} )^2 &= \frac{(\mathbbm{h}_\mathrm{a} \cdot \boldsymbol{\mathsf{F}} \boldsymbol a^{\mathrm u})^2}{|\boldsymbol{\mathsf{F}} \boldsymbol a^{\mathrm u}|^2} =
\frac{ (\mathbbm{h}_\mathrm{a} \cdot\boldsymbol d_3)^2 + 2 X_\alpha (\mathbbm{h}_\mathrm{a} \cdot \boldsymbol{\kappa} \times \boldsymbol{d}_\alpha)  (\mathbbm{h}_\mathrm{a} \cdot\boldsymbol d_3)+O(\varepsilon^2)}
{1 + 2 X_\alpha \boldsymbol{d}_3 \cdot \boldsymbol{\kappa} \times \boldsymbol{d}_\alpha + O(\varepsilon^2)}  
\\
&=\frac{ (\mathbbm{h}_\mathrm{a} \cdot\boldsymbol d_3)^2 (1 - 2 X_\alpha \boldsymbol{d}_3 \cdot \boldsymbol{\kappa} \times \boldsymbol{d}_\alpha ) + 2 X_\alpha (\mathbbm{h}_\mathrm{a} \cdot \boldsymbol{\kappa} \times \boldsymbol{d}_\alpha)  (\mathbbm{h}_\mathrm{a}\cdot\boldsymbol d_3) + O(\varepsilon^2)}{1-O(\varepsilon^2)}
\\
&=(\mathbbm{h}_\mathrm{a} \cdot\boldsymbol d_3)^2 -2 X_\alpha (\boldsymbol{\kappa} \cdot \boldsymbol{d}_\alpha \times \boldsymbol{d}_\beta ) (\mathbbm{h}_\mathrm{a} \cdot\boldsymbol{d}_\alpha) (\mathbbm{h}_\mathrm{a} \cdot\boldsymbol{d}_\beta) + O(\varepsilon^2) \,.
\end{split}
\end{equation} 
Let $\mathfrak C^{\rm u}(s)$ denote the part of the rod delimited by the sections $\mathscr{S} (0)$ and $\mathscr{S} (s)$, and let $J$ be the Jacobian of the map $\hat{\boldsymbol x}:\mathcal C\to\mathfrak C^{\rm u}$, 
\begin{equation}  \label{kth24}   
J = \boldsymbol{g}_1 \times \boldsymbol{g}_2 \cdot \boldsymbol{g}_3 = 1 + \boldsymbol{d}^{\mathrm u}_3 \cdot X_\alpha \boldsymbol{\kappa}^\mathrm{u} \times \boldsymbol{d}^\mathrm{u}_\alpha \,;    
\end{equation} 
neglecting the constant term in the expression \eqref{kth20b} of $\hat{\psi}_m$, the magnetostatic energy of $\mathfrak C^{\rm u}(s)$ is 
\begin{equation} \label{kth25} \nonumber 
\begin{aligned}
\Psi_m (s) &=  \int_{\mathfrak C^{\rm u}(s)} \hat{\psi}_m dV = - \frac{1}{2} \mu_0\nu V_\Pi \chi  \int_{\mathfrak C^{\rm u}(s)} \frac{(\mathbbm{h}_\mathrm{a} \cdot \boldsymbol{\mathsf{F}} \boldsymbol a^{\mathrm u})^2}{|\boldsymbol{\mathsf{F}} \boldsymbol a^{\mathrm u}|^2} dV 
\\ 
&= - \frac{1}{2} \mu_0\nu V_\Pi \chi  \int_{\Sigma\times(0,s)}\frac{(\mathbbm{h}_\mathrm{a} \cdot \boldsymbol{\mathsf{F}} \boldsymbol a^{\mathrm u})^2}{|\boldsymbol{\mathsf{F}} \boldsymbol a^{\mathrm u}|^2}  \, J \, dX_1 dX_2 d \varsigma
\\ 
&= - \frac{1}{2} \mu_0\nu V_\Pi \chi \int_0^s \int_\Sigma \Big( (\mathbbm{h}_\mathrm{a} \cdot\boldsymbol d_3)^2 (1 + X_\alpha \, \boldsymbol{d}^{\mathrm u}_3 \cdot \boldsymbol{\kappa}^\mathrm{u} \times \boldsymbol{d}^\mathrm{u}_\alpha) 
\\
& \quad - 2 X_\alpha (\boldsymbol{\kappa} \cdot \boldsymbol{d}_\alpha \times \boldsymbol{d}_\beta ) (\mathbbm{h}_\mathrm{a} \cdot\boldsymbol{d}_\alpha) (\mathbbm{h}_\mathrm{a} \cdot\boldsymbol{d}_\beta) \Big) dX_1 dX_2 d \varsigma + O(\varepsilon^2) \,.  
\end{aligned}
\end{equation}
Since the centroid of $\Sigma$ belongs to the axes $X_1$ and $X_2$, we have  
\begin{equation}  \label{kth26}  
\int_\Sigma X_\alpha dX_1 dX_2 = 0 \,, \qquad \quad \alpha = 1, 2 \,, 
\end{equation}
and we conclude that, within the same order of approximation at which the Kirchhoff's theory holds, the magnetostatic energy per unit length along $\mathscr{C}^{\mathrm u}$ is
\begin{equation}  \label{kth27}  
\psi_m = \frac{d \Psi_m}{ds} = - \frac{1}{2} A\mu_0\nu V_\Pi \chi (\mathbbm{h}_\mathrm{a} \cdot\boldsymbol d_3)^2  = - \frac{1}{2} \tilde{\eta} \, (\boldsymbol{e} \cdot \boldsymbol{d}_3)^2  \,,
\end{equation}
where $A$ is the area of the cross sections $\mathscr{S}$ and $\boldsymbol{e}$ is a unit vector parallel to $\mathbbm{h}_\mathrm{a}$, and where we have put 
\begin{equation}  \label{kth28}
\tilde{\eta} = A\mu_0\nu V_\Pi \chi | \mathbbm{h}_\mathrm{a} |^2 \,.  
\end{equation}
The constant $\tilde{\eta}$ has the dimensions of a force times the square of a length, divided by the square of a current. 

\subsection{Equilibrium of magneto-elastic rods} 

We deduce the equilibrium equations of a magneto-elastic rod, subject to a uniform magnetic field and to forces and couples applied at the ends, by taking the variation of its elastic and magnetic energy under the condition that axial inextensibility be preserved.   
A variational derivation of the equilibrium equations of Kirchhoff's rods in a purely mechanical context, for deformations at the level of generality of those considered Section \ref{S:Ktr}, has been presented in \citet{ML2003}. 
To deduce the equilibrium equations of a magneto-elastic rod, we employ the results of that paper, adding to the variation of the mechanical quantities the term arising from the variation of the magnetic energy. 

The small displacement $\bar{\boldsymbol{u}}$ does not affect the equilibrium equations \eqref{kth16} and the constitutive equation \eqref{kth17};  once the curvature vector $\boldsymbol{\kappa}^\mathrm{u}$ has been determined,  $\bar{\boldsymbol{u}}$ is obtained by integrating the three-dimensional equilibrium equations with the appropriate conditions which hold at the lateral surface of the rod (cf. \citet{AL1944}, Sections 257-258;  \citet{ED1992}, Sect.3). 
Thus, a configuration $\mathfrak{C}$ of a rod is determined when the displacement of the points of the axial curve,
\begin{equation}  \label{elmr1}  \nonumber 
\boldsymbol{u} (s) = \boldsymbol{x} (s) - \boldsymbol{x}^\mathrm{u} (s) \,, 
\end{equation}
and the rotation $\boldsymbol{Q}$ defined by Equation \eqref{kth14} are known. 
As a consequence of the assumption of inextensibility, it is 
\begin{equation}  \label{elmr2}
\boldsymbol{u}^\prime (s) = (\boldsymbol{Q} (s) - \boldsymbol{I}) \boldsymbol{d}^\mathrm{u}_3 (s)  \,;   
\end{equation}
conversely, taking the definition of $\boldsymbol{u}$ into account, \eqref{elmr2} yields $\boldsymbol{x}^\prime (s) = \boldsymbol{Q} (s) \boldsymbol{d}^\mathrm{u}_3 (s)$ which implies the inextensibility condition \eqref{kthr3}. 
Hence, equation \eqref{elmr2} can be regarded as an expression of the condition that the rod axis is inextensible. 

Let $\delta \boldsymbol{u}$ and $\delta \boldsymbol{Q}$ denote variations of $\boldsymbol{u}$ and $\boldsymbol{Q}$, respectively, and let $\delta \boldsymbol{q}$ denote the vector associated with the skew tensor $(\delta \boldsymbol{Q}) \boldsymbol{Q}^\mathsf{T}$; then, the variation of the vectors $\boldsymbol{d}_i$ are   
\begin{equation}  \label{elmr3}
\delta \boldsymbol{d}_i = \delta ( \boldsymbol{Q} \boldsymbol{d}^\mathrm{u}_i ) = (\delta \boldsymbol{Q}) \boldsymbol{Q}^\mathsf{T} \boldsymbol{d}_i = \delta \boldsymbol{q} \times \boldsymbol{d}_i \,, \qquad i = 1, 2, 3 \,. 
\end{equation}
Making use of \eqref{kthr4}, \eqref{kth14}, \eqref{kth15}, \eqref{kth17}, \eqref{kth18}, and \eqref{elmr3}, it can be shown that 
\begin{equation}  \label{elmr4} \nonumber 
\delta \boldsymbol{\mu} = \delta \boldsymbol{q}^\prime + \delta \boldsymbol{q} \times \boldsymbol{\mu} \,, 
\qquad \quad 
\delta \boldsymbol{M} = \boldsymbol{C} \delta \boldsymbol{q}^\prime + \delta \boldsymbol{q} \times \boldsymbol{M} \,; 
\end{equation}
these equations imply that the variation of the strain energy density is 
\begin{equation}  \label{elmr5}
\delta \psi_e = (\boldsymbol{M} \cdot \delta \boldsymbol{q})^\prime - \boldsymbol{M}^\prime \cdot \delta \boldsymbol{q} \,. 
\end{equation}
The constraint \eqref{elmr2} can be taken into account through the integral
\begin{equation}  \label{elmr6}  \nonumber 
\mathscr{L} = \int_0^\ell \boldsymbol{\Phi} \cdot \big( \boldsymbol{u}^\prime - (\boldsymbol{Q} - \boldsymbol{I} ) \boldsymbol{d}_3^\mathrm{u} \big) \, ds \,,  
\end{equation}
where the vector $\boldsymbol{\Phi}$ is a Lagrange multiplier; the variation of $\mathscr{L}$ furnishes  
\begin{equation}  \label{elmr7}
\delta \mathscr{L} = - \int_0^\ell \big( \boldsymbol{\Phi}^\prime \cdot \delta \boldsymbol{u} - \boldsymbol{d}_3 \times \boldsymbol{\Phi} \cdot \delta \boldsymbol{q} \big) ds + \left[ \boldsymbol{\Phi} \cdot \delta \boldsymbol{u} \right]_0^\ell \,. 
\end{equation}
Since the quantity $\tilde{\eta}$ defined by \eqref{kth28} is a constant, it is easily seen that the variation of the magnetic energy density \eqref{kth27} is 
\begin{equation}  \label{elmr8}
\delta \psi_m = - \frac{1}{2} \tilde{\eta} \,\delta (\boldsymbol{e} \cdot \boldsymbol{d}_3)^2 = - \tilde{\eta} \, (\boldsymbol{d}_3 \cdot \boldsymbol{e} ) \, \boldsymbol{d}_3 \times \boldsymbol{e} \cdot \delta \boldsymbol{q} \,.  
\end{equation}
Finally, we collect the results \eqref{elmr5}, \eqref{elmr7}, and \eqref{elmr8}, and impose that the variation of the magneto-elastic energy $\mathscr{E}$ of the rod subject to the constraint \eqref{elmr2} vanish:
\begin{equation}  \label{elmr9} \nonumber 
\begin{split} 
\delta ( \mathscr{E} + \mathscr{L}) &= \delta \int_0^\ell ( \psi_e + \psi_m ) ds + \delta \mathscr{L} = 
\\ 
&= - \int_0^\ell \big( (\boldsymbol{M}^\prime + \boldsymbol{d}_3 \times \boldsymbol{\Phi} + (\boldsymbol{e} \cdot \boldsymbol{d}_3) \, \boldsymbol{e} \times \boldsymbol{d}_3 ) \cdot \delta \boldsymbol{q} + \boldsymbol{\Phi}^\prime \cdot \delta \boldsymbol{u} \big) ds 
\\
&\quad + \left[ \boldsymbol{M} \cdot \delta \boldsymbol{q} + \boldsymbol{\Phi} \cdot \delta \boldsymbol{u} \right]_0^\ell = 0 \,. 
\end{split} 
\end{equation}
When the multiplier $\boldsymbol{\Phi}$ is identified with the resultant force $\boldsymbol{F}$, this equation yields that the equilibrium equations of a Kirchhoff's rod subject to end loads and a distribution of magnetic couples per unit length of the axial curve, are: 
\begin{equation} \label{elmr10} 
\boldsymbol{F}^\prime  = \boldsymbol{0} \,,
\qquad \quad 
\boldsymbol{M}^\prime + \boldsymbol{d}_3 \times \boldsymbol{F} + \tilde{\eta} \, (\boldsymbol{d}_3 \cdot \boldsymbol{e} ) \, \boldsymbol{d}_3 \times \boldsymbol{e} = \boldsymbol{0} \,. 
\end{equation}
Clearly, these equations coincide with equations \eqref{kth16} in which the force per unit length vanishes, $\boldsymbol{f} = \boldsymbol{0}$, and the couple per unit length is due to the magnetic interactions, $\boldsymbol{m} = \tilde{\eta} \, (\boldsymbol{d}_3 \cdot \boldsymbol{e} ) \, \boldsymbol{d}_3 \times \boldsymbol{e}$. 
As shown by equations \eqref{kth28} and \eqref{elmr10}, only the direction of the field $\mathbbm{h}_{\rm a}$ is relevant to the equilibrium of the rod, while its orientation is immaterial. 

\remark 
The magnetic couple per unit length of the axial curve and the equilibrium equations of a magneto-elastic rod can be obtained through an alternative procedure that employs the density of magnetization for the problem under consideration in the general formula giving the magnetic couple produced by a magnetic field on a magnetized body. 

The couple per unit referential volume $\hat{\boldsymbol{l}}$ exerted by the magnetic induction field $\mathbbm{b}$ on the magnetization $\mathbbm{M}$ is $\hat{\boldsymbol{l}} = \mathbbm{M} \times \mathbbm{b}$ \citep[cf., e.g.,][Sect.8.3]{HT1990}; by putting  $\mathbbm{b} = \mu_0 \mathbbm{h}_{\rm a}$ and $\mathbbm{M} = \nu V_\Pi \mathbbm{m}$, with $\mathbbm{m}$ given by 
\eqref{elmr11}, it is 
\begin{equation} \label{elmr12}  
\hat{\boldsymbol{l}} = \mathbbm{M} \times \mathbbm{b} =  \nu V_\Pi \mathbbm{m} \times \mu_0 \mathbbm{h}_{\rm a} =  \mu_0 \nu V_\Pi \chi ( \mathbbm{h}_\mathrm{a} \cdot \boldsymbol{a} ) \boldsymbol{a} \times  \mathbbm{h}_{\rm a} \,.  
\end{equation}
In a deformation of a Kirchhoff's rod in which the magnetic particles are parallel to the longitudinal material fibers, $\boldsymbol{a} = \boldsymbol{\mathsf{F}} \boldsymbol{g}_3 / |\boldsymbol{\mathsf{F}} \boldsymbol{g}_3|$, and $\boldsymbol{\mathsf{F}} \boldsymbol{g}_3 = (\boldsymbol{d}_3 + \boldsymbol{\kappa} \times X_\alpha \boldsymbol{d}_\alpha) + O(\varepsilon^2)$; hence,  
\begin{equation} \label{elmr13} 
\begin{split} 
&( \mathbbm{h}_\mathrm{a} \cdot \boldsymbol{a} ) \boldsymbol{a} \times  \mathbbm{h}_{\rm a} = 
\\
&= \frac{(\boldsymbol{d}_3 \cdot \mathbbm{h}_\mathrm{a}) \big( \boldsymbol{d}_3 \times  \mathbbm{h}_\mathrm{a} + X_\alpha (\boldsymbol{\kappa} \times \boldsymbol{d}_\alpha) \times \mathbbm{h}_\mathrm{a} \big) +  X_\alpha \mathbbm{h}_\mathrm{a} \cdot \boldsymbol{\kappa} \times \boldsymbol{d}_\alpha (\boldsymbol{d}_3 \times  \mathbbm{h}_\mathrm{a}) + O(\varepsilon^2) }{ 1 + 2 \, X_\alpha \boldsymbol{d}_3 \cdot \boldsymbol{\kappa} \times \boldsymbol{d}_\alpha  + O(\varepsilon^2)}  
\\ 
&= (\boldsymbol{d}_3 \cdot \mathbbm{h}_\mathrm{a}) \boldsymbol{d}_3 \times \mathbbm{h}_\mathrm{a} +  X_\alpha \big( (\boldsymbol{d}_3 \cdot \mathbbm{h}_\mathrm{a}) (\boldsymbol{\kappa} \times  \boldsymbol{d}_\alpha) \times \mathbbm{h}_\mathrm{a} + (\boldsymbol{d}_3 \times \mathbbm{h}_\mathrm{a}) \mathbbm{h}_\mathrm{a} \cdot \boldsymbol{\kappa} \times  \boldsymbol{d}_\alpha 
\\ 
&\qquad \qquad \qquad \qquad \qquad \qquad - 2 (\boldsymbol{d}_3 \cdot \mathbbm{h}_\mathrm{a}) (\boldsymbol{d}_3 \times \mathbbm{h}_\mathrm{a}) \boldsymbol{d}_3 \cdot  \boldsymbol{\kappa} \times \boldsymbol{d}_\alpha \big) + O(\varepsilon^2) \,. 
\end{split} 
\end{equation}
Taking into account \eqref{kth24} and \eqref{kth26}, equations \eqref{elmr12} and \eqref{elmr13} imply that the couple acting on the part $\mathfrak{C}^{\rm u}(s)$ of the rod is 
\begin{equation} \label{elmr14} \nonumber   
\begin{aligned}
\boldsymbol{L}(s) &= \int_{\mathfrak C^{\rm u}(s)}  \hat{\boldsymbol{l}}  \, dV = \mu_0 \nu V_\Pi \chi \int_{\Sigma\times(0, s)}  ( \mathbbm{h}_\mathrm{a} \cdot \boldsymbol{a} ) \boldsymbol{a} \times  \mathbbm{h}_{\rm a}  \, J \, dX_1 dX_2 d\varsigma
\\
&=A \mu_0 \nu V_\Pi \chi \int_0^s (\boldsymbol{d}_3 \cdot \mathbbm{h}_\mathrm{a}) \boldsymbol{d}_3 \times \mathbbm{h}_\mathrm{a} \, d\varsigma + O(\varepsilon^2) 
\\
&= \tilde{\eta} \int_0^s \, (\boldsymbol{d}_3 \cdot \boldsymbol{e} ) \, \boldsymbol{d}_3 \times \boldsymbol{e} \, d \varsigma + O(\varepsilon^2) \,. 
\end{aligned}
\end{equation}
It follows that, when terms of order $O(\varepsilon^2)$ are neglected, the magnetic couple $\boldsymbol{l}$ per unit length of the axial curve is 
\begin{equation} \label{elmr15} \nonumber  
\boldsymbol{l} = \frac{d \boldsymbol{L}}{ds} = \tilde{\eta} \, (\boldsymbol{d}_3 \cdot \boldsymbol{e} ) \, \boldsymbol{d}_3 \times \boldsymbol{e} \,; 
\end{equation}
then, the equilibrium equations \eqref{elmr10} for a magneto-elastic rod are obtained by putting $\boldsymbol{f} = \boldsymbol{0}$ and $\boldsymbol{m} = \boldsymbol{l} = \tilde{\eta} \, (\boldsymbol{d}_3 \cdot \boldsymbol{e} ) \, \boldsymbol{d}_3 \times \boldsymbol{e}$ into the usual equilibrium equations \eqref{kth16} of Kirchhoff's theory.

\section{Exact equilibrium solutions in terms of elliptic functions}

\begin{figure}[h!]   
\begin{center}
\begin{tikzpicture}[>=angle 90,scale=1.0]]
\tikzset{ground2/.style={fill,pattern=north east lines,draw=none, minimum width=0.05cm,minimum height=0.05cm}}
\tikzset{ground/.style={fill,pattern=north east lines,draw=none, minimum width=0.75cm, minimum height=0.1cm}}
\tikzset{groundv/.style={fill,pattern=north east lines,draw=none, minimum width=0.05cm,minimum height=0.75cm}}
\tikzset{groundv2/.style={fill,pattern=north east lines,draw=none, minimum width=0.05cm,minimum height=0.04cm}}
\pgfmathsetmacro{\lun}{6}
\coordinate (A) at (0,0);
\coordinate (B) at ($(A)+(\lun,0)$);
%
%
\draw[black] (A) -- (B); 
%
%
\node (h1) at (A) [groundv,anchor=east]{};
\draw (h1.north east) -- (h1.south east);
%
%
\coordinate (O) at (0,0);
\draw [black,->] (O) -- ++(20:1.2) node[anchor=south] {$z$};
\draw [black,->] (O)  -- ++(-70:1.2) node[anchor=west] {$y$};
%
%
\node [black] at ($(A)+(-0.1,0.7)$) {$s=0$};
\node [black] at ($(B)+(-0.1,0.3)$) {$s=\ell$};
%
%
\draw [black, thick, ->] ($(B)+(0.5,0.25)$)-- ++(45:1.5);
\draw [black] ($(B)+(1.1,1.0)$) node[left] {$\mathbbm{h}_\mathrm{a}$}; 
\draw [black, thick, ->] ($(B)+(1.75,0.25)$)-- ++(45:0.5);
\draw [black] ($(B)+(1.9,0.6)$) node[left] {$\boldsymbol{e}$};
%
%
\draw [black, thick, ->] ($(B)+(0.02,-0.02)$)--node[right] {$\boldsymbol{F}$} ++(-70:1.5);
%
%
\draw[black] plot [domain=0:\lun] (\x, {(-\x/\lun*\x/\lun-0.5*\x/\lun*\x/\lun*\x/\lun)}); 
%
%
\coordinate (P) at (4*\lun/5,-16/25-0.5*64/125);
\draw[black,fill] (P) circle (0.3mm);
\draw [black, thick, ->] (P) -- ++(0.8,-0.35); 
\node [black] at ($(P)+(0.55,0.1)$) {$\boldsymbol{d}_3$};
\draw [black] (P) -- ++(-70:1); 
\draw[black,->] ($(P)+(-70:0.5)$) arc (-70:-23:0.5); 
\node[black] at ($(P)+(-0.01,-0.6)$) {$\vartheta$};
\draw[black,->] ($(A)+(-70:0.5)$) arc (-70:0:0.5); 
\node[black] at ($(A)+(0.6,-0.45)$) {$\vartheta_\mathrm{o}$};
\end{tikzpicture}
\end{center}
\vspace{-5pt}
\caption{Cantilever immersed in a uniform magnetic field  $\mathbbm{h}_\mathrm{a}$ and subject to a force $\boldsymbol{F}$ applied at the free end; the unit vector $\boldsymbol{e}$ is such that $\boldsymbol{e} \times \mathbbm{h}_\mathrm{a} = \boldsymbol{0}$.}  \label{Fi:figura1}
\end{figure}
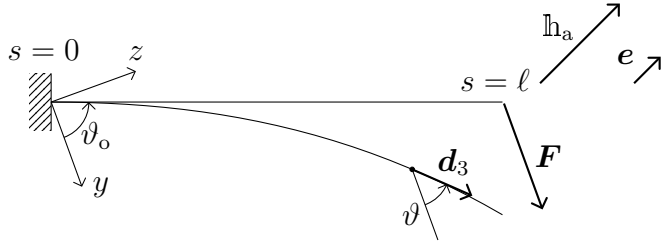

In this Section we present some examples of exact solutions of the equilibrium equations \eqref{elmr10} obtained by means of the Weierstrass elliptic functions. 
We consider the plane problem of a cantilever, prismatic and  twist-free in the configuration $\mathfrak{C}^\mathrm{u}$, subject to the action of a magnetic field $\mathbbm{h}_\mathrm{a}$ and a force $\boldsymbol{F}$ applied at its free end.  
We employ a Cartesian coordinate system $(x, y, z)$, whose base vectors are $(\boldsymbol{e}_x, \boldsymbol{e}_y, \boldsymbol{e}_z)$, with origin on the rod axis at the fixed end, the $y$-axis having the oriented direction of $\boldsymbol{F}$; 
as in Section \ref{S:eKr}, the unit vector $\boldsymbol{e}$ gives the direction of the magnetic field (Figure \ref{Fi:figura1}).  
The fixed end of the rod corresponds to $s=0$; the angle from $y$ to $\boldsymbol{d}_3$ is denoted by $\vartheta$, and its value at $s=0$ is $\vartheta_\mathrm{o}$. 

\smallskip 

\noindent
The boundary conditions for the equilibrium problem of the rod are
\begin{equation} \label{exsl1}
\boldsymbol{x} (0) = \boldsymbol{0} \,, \qquad \quad \boldsymbol{M} (\ell) = \boldsymbol{0} \,. 
\end{equation}
Since the rod is assumed to be prismatic and twist-free in $\mathfrak{C}^\mathrm{u}$, the curvature vector in that configuration vanishes, $\boldsymbol{\kappa}^\mathrm{u} = \boldsymbol{0}$; then, the measure of deformation \eqref{kth15} reduces to $\boldsymbol{\mu} = \boldsymbol{\kappa}$ and, making use of \eqref{kthr5}$_1$, the constitutive equation \eqref{kth17} becomes 
\begin{equation} \label{exsl2}
\boldsymbol{M} =  \boldsymbol{C} \boldsymbol{\kappa} = EI \boldsymbol{d}_3 \times \boldsymbol{d}_3^{\, \prime} + GJ \kappa_3 \boldsymbol{d}_3 \,. 
\end{equation}
Equilibrium equation \eqref{elmr10}$_1$ requires that the resultant force $\boldsymbol{F}$ be constant.  
In view of equation \eqref{exsl2}, the scalar product of \eqref{elmr10}$_2$ with $\boldsymbol{d}_3$ furnishes $\boldsymbol{M}^\prime \cdot \boldsymbol{d}_3 = (\boldsymbol{M} \cdot \boldsymbol{d}_3)^\prime = 0$, that implies $\kappa_3^\prime = 0$; thus, the twist is uniform along the axis as in rods subject to terminal loads only. 
In particular, in the present case, the condition \eqref{exsl1}$_2$ specifies that the twist in the equilibrium configuration $\mathfrak{C}$ vanishes. 

In the study of exact solutions for the system \eqref{elmr10}, \eqref{exsl1}, and \eqref{exsl2}, we consider deformations of the rod occurring in the plane $(y, z)$, for which we have:  
\begin{equation} \label{exsl3} \nonumber 
\boldsymbol{F} = |\boldsymbol{F}| \boldsymbol{e}_y \,, \qquad \mathbbm{h}_\mathrm{a} = \pm |\mathbbm{h}_\mathrm{a}| \boldsymbol{e} \,, \qquad \boldsymbol{d}_3 = y^\prime \boldsymbol{e}_y + z^\prime \boldsymbol{e}_z \,; 
\end{equation}
in components the condition of inextensibility \eqref{kthr3} is  
\begin{equation} \label{exsl4} 
(y^\prime)^2 + (z^\prime)^2 = 1 \,. 
\end{equation}
By putting 
\begin{equation} \label{exsl5} \nonumber  
\beta = \frac{\boldsymbol{F} \cdot \boldsymbol{e}_y}{EI} > 0 \,, \qquad \quad \eta = \frac{\tilde{\eta}}{EI} = \frac{1}{EI} A\mu_0\nu V_\Pi \chi |\mathbbm{h}_\mathrm{a}|^2 > 0 \,,  
\end{equation}
the equilibrium equation \eqref{elmr10}$_2$ can be written in the form   
\begin{equation} \label{exsl6}
\big( \boldsymbol{d}_3 \times \boldsymbol{d}_3^{\, \prime} \big)^\prime + \beta \boldsymbol{d}_3 \times \boldsymbol{e}_y + \eta \, (\boldsymbol{d}_3 \cdot \boldsymbol{e} ) \, \boldsymbol{d}_3 \times \boldsymbol{e} = \boldsymbol{0} \,.   
\end{equation}
Henceforth we use dimensionless variables obtained by taking $\sqrt{I/A}$ as unit of length, $EA$ as unit of force, and $\sqrt{EA/ \mu_0}$ as unit of electric current; denoting by a bar the new variables, we have  
\begin{equation} \label{exsl7}  \nonumber 
\begin{split} 
\{\overline{s}, \overline{y},  \overline{z} \} &= \{s, y, z\} \sqrt{\frac{A}{I}}  \,, \qquad 
\overline{k} = k \sqrt{\frac{I}{A}} \,, \qquad \overline{F} = \frac{F}{EA} = \beta \frac{I}{A} = \overline{\beta}  \,, 
\\  
\overline{\mathbbm{h}}_\mathrm{a} &= \mathbbm{h}_\mathrm{a} \sqrt{\frac{I \mu_0}{E A^2}} \,, \qquad \overline{\eta} = \eta \frac{I}{A}  \,;
\end{split} 
\end{equation}
after substitution in the previous equations, we omit the bars so that the formal aspect of the equations is unaltered.   

To arrive at a differential equation of the type soluble by means of elliptic functions, we take the scalar product of equation \eqref{exsl6} with $\boldsymbol{d}_3 \times \boldsymbol{d}_3^{\, \prime}$; after integration of the result, we have 
\begin{equation} \label{exsl8} 
\kappa_x^2 = \gamma - 2 \beta y^\prime - \eta (\boldsymbol{d}_3 \cdot \boldsymbol{e})^2 \,,   
\end{equation}
where 
\begin{equation} \label{exsl9} \nonumber   
\kappa_x = \boldsymbol{d}_3 \times \boldsymbol{d}_3^{\, \prime} \cdot \boldsymbol{e}_x = k \boldsymbol{b} \cdot \boldsymbol{e}_x = \pm k \,, 
\end{equation}
and $\gamma$ is an integration constant.  
The component along $y$ of the equation $\boldsymbol{d}_3^\prime = \boldsymbol{\kappa} \times \boldsymbol{d}_3$ is 
\begin{equation} \label{exsl9a} 
y^{\prime \prime} = - \kappa_x z^\prime \,; 
\end{equation}
the square of this equation and the inextensibility condition \eqref{exsl4} yield   
\begin{equation} \label{exsl10} \nonumber  
(y^{\prime \prime})^2 = \kappa_x^2 \big( 1- (y^\prime)^2 \big) \,;  
\end{equation}
finally, substituting $\kappa_x^2$ from \eqref{exsl8}, we obtain   
\begin{equation} \label{exsl11} 
(y^{\prime \prime})^2 = \big(\gamma - 2 \beta y^\prime - \eta (\boldsymbol{d}_3 \cdot \boldsymbol{e})^2 \big) \big( 1- (y^\prime)^2 \big) \,. 
\end{equation}
The form of this equation shows that it can be solved by means of elliptic function when $\boldsymbol{e} = \boldsymbol{e}_y$ so that $(\boldsymbol{d}_3 \cdot \boldsymbol{e})^2 = (y^\prime)^2$, and when $\boldsymbol{e} = \boldsymbol{e}_z$ so that $(\boldsymbol{d}_3 \cdot \boldsymbol{e})^2 = 1- (y^\prime)^2$. 
In the first case the magnetic field $\mathbbm{h}_\mathrm{a}$ is parallel, in the second case orthogonal, to the force $\boldsymbol{F}$; in both cases \eqref{exsl11} is an equation of the type 
\begin{equation} \label{exsl12} 
(\xi^\prime )^2 = f (\xi) = \hat{f} (\xi) ( 1- \xi^2) \,, 
\end{equation}
where 
\begin{equation} \label{exsl12a}  \nonumber 
\xi = y^\prime = \cos \vartheta \,, 
\end{equation}
and $f (\xi)$ and $\hat{f} (\xi)$ are a polynomials in $\xi$ of the fourth and second degree, respectively. 

\remark 
For the case in which $\mathbbm{h}_\mathrm{a}$ is parallel to $\boldsymbol{F}$, it is possible to show that when the condition \eqref{exsl1}$_2$ holds, the rod undergoes a planar deformation. 
In fact, denoted by $\boldsymbol{e}$ the direction of $\mathbbm{h}_\mathrm{a}$ and $\boldsymbol{F}$, the equilibrium equation \eqref{elmr10}$_2$ can be written as 
\begin{equation} \label{exslrm1}
\boldsymbol{M}^\prime + \big( | \boldsymbol{F} | + \tilde{\eta} \, (\boldsymbol{d}_3 \cdot \boldsymbol{e} ) \big) \boldsymbol{d}_3 \times \boldsymbol{e} = \boldsymbol{0} \,.    
\end{equation}
The discussion following Equation \eqref{exsl2} has shown that, in the considered problem, $\kappa_3 = 0$. 
Taking into account that $\boldsymbol{e}$ is a constant vector, from the scalar product of \eqref{exslrm1} with $\boldsymbol{e}$ one has 
\begin{equation} \label{exslrm3} \nonumber 
\boldsymbol{M}^\prime \cdot \boldsymbol{e} = (\boldsymbol{M} \cdot \boldsymbol{e})^\prime = 0 \,; 
\end{equation}
this equation and the condition \eqref{exsl1}$_2$ show that $\boldsymbol{M} \cdot \boldsymbol{e} = 0$; thus, since $\kappa_3=0$, equation \eqref{exsl2} yields 
\begin{equation} \label{exslrm4} 
\frac{1}{EI} \, \boldsymbol{M} \cdot \boldsymbol{e} = \boldsymbol{d}_3 \times \boldsymbol{d}_3^\prime\cdot \boldsymbol{e} = 0 \,. 
\end{equation}
It follows from equations \eqref{exsl2}, \eqref{exslrm1} and \eqref{exslrm4} that    
\begin{equation} \label{exslrm5}
\begin{split} 
\boldsymbol{M}^\prime \cdot \boldsymbol{d}_3^\prime = EI ( \boldsymbol{d}_3 \times \boldsymbol{d}_3^\prime)^\prime \cdot \boldsymbol{d}_3^\prime &= - \big( | \boldsymbol{F} | + \tilde{\eta} \, (\boldsymbol{d}_3 \cdot \boldsymbol{e} ) \big) \boldsymbol{d}_3 \times \boldsymbol{e} \cdot \boldsymbol{d}_3^\prime 
\\
&= \big( | \boldsymbol{F} | + \tilde{\eta} \, (\boldsymbol{d}_3 \cdot \boldsymbol{e} ) \big) \boldsymbol{d}_3 \times \boldsymbol{d}_3^\prime \cdot \boldsymbol{e} = 0 \,;   
\end{split}   
\end{equation}
moreover, by use of the Serret-Frenet formulae one has    
\begin{equation} \label{exslrm6}
\tau k^2 = \boldsymbol{d}_3 \times \boldsymbol{d}_3^\prime \cdot \boldsymbol{d}_3^{\prime \prime} = - ( \boldsymbol{d}_3 \times \boldsymbol{d}_3^\prime)^\prime \cdot \boldsymbol{d}_3^\prime  \,.  
\end{equation}
Equations \eqref{exslrm5} and \eqref{exslrm6} imply that, under the present assumptions, $\tau k^2 = 0$; hence, when $k$ vanishes at most at isolated points, $\tau = 0$, that is, the axial curve is planar.

\subsection{Solution for $\mathbbm{h}_\mathrm{a}$ parallel to $\boldsymbol{F}$} 
When $\mathbbm{h}_\mathrm{a}$ is parallel to $\boldsymbol{F}$, the polynomials $f (\xi)$ and $\hat{f} (\xi)$ are 
\begin{equation} \label{exsl13}  \nonumber 
f(\xi) = \eta \xi^4 + 2 \beta \xi^3 - (\gamma + \eta) \xi^2 - 2 \beta \xi + \gamma \,, \qquad \hat{f} (\xi) = - \eta \xi^2 - 2 \beta \xi + \gamma \,.  
\end{equation}
Because $\xi$ is equal to $\cos  \vartheta$, its values must be in the interval $[-1,1]$. 
The roots $\xi_a$ and $\xi_b$ of $\hat{f} (\xi)$, with $\xi_a < \xi_b$, are 
\begin{equation}  \label{exsl14} 
\left. \begin{matrix}
\xi_a \\ \xi_b  
\end{matrix} \right\} 
= \frac{\beta \mp \sqrt{\beta^2 + \gamma \eta}}{- \eta} = - \frac{\beta}{\eta} \Big( 1 \pm \sqrt{1 + \frac{\gamma \eta}{\beta^2}} \Big)  \,.   
\end{equation} 
Equation \eqref{exsl12} requires $\hat{f} (\xi) \ge 0$; since the coefficient of $\xi^2$ in $\hat{f} (\xi)$ is negative, $\hat{f} (\xi)$ is nonnegative for $\xi$ in the interval whose ends are the roots of $\hat{f} (\xi)$, and the admissible values of $\xi$ are those which satisfy the relations   
\begin{equation}  \label{exsl15} 
\xi \in [ -1, 1 ] \,, \qquad \xi \in \, [ \xi_a, \xi_b ] \,. 
\end{equation} 
To solve equation \eqref{exsl12} we consider $\xi$ as a function of a complex variable $u$ (we use the same symbol for $\xi$ as function of $u$ and $s$, but it should be clear from the context which is the function involved)  and, denoted by $a$ a root of $f(\xi)$, we put  \citep[cf.][Sect.124]{LB1930}   
\begin{equation} \label{exsl16} 
\xi(u) = a + \frac{b}{\varkappa(u) - c} \,,  
\end{equation}
where    
\begin{equation} \label{exsl17} \nonumber  
b = \frac{f^\prime(a)}{4} \,, \qquad \quad c = \frac{f^{\prime \prime} (a)}{24} \,;  
\end{equation}
by choosing $a=1$, we have 
\begin{equation} \label{exsl18} 
b = \beta + \frac{\eta - \gamma}{2} \,, \qquad \quad c = \frac{\beta + \eta}{2} - \frac{\gamma +\eta}{12} \,.  
\end{equation} 
In the interval of values of interest for the problem under consideration, we write the variable $u$ as $u = u_\mathrm{o} + s$, with $u_\mathrm{o}$ a complex constant to be determined.  
Then, by introducing \eqref{exsl16} into equation \eqref{exsl12}, we have    
\begin{equation} \label{exsl19} 
\left( \xi^\prime \right)^2 = \left( \frac{d \xi}{d \varkappa} \frac{d \varkappa}{du} \frac{d u}{ds} \right)^2 = \left( \frac{-b}{(\varkappa - c )^2} \right)^2 \big( 4 \varkappa^3 - g_2 \varkappa - g_3 \big) \,, 
\end{equation} 
which shows that $\varkappa$ is the Weierstrass function $\wp$,   
\begin{equation} \label{exsl19a} \nonumber  
\varkappa (u) = \wp (u; g_2, g_3) = \wp (u) \,, 
\end{equation} 
or, in the interval of values of $u$ corresponding to the points of the rod axis,  
\begin{equation} \label{exsl20} \nonumber  
\varkappa (u_\mathrm{o} + s) = \wp (u_\mathrm{o} + s; g_2, g_3) = \wp (u_\mathrm{o} + s) \,, \qquad \quad s\in (0, \ell) \,, 
\end{equation} 
with $u_\mathrm{o}$ the value of $u$ at the fixed end of the rod;  
hence, the function $\xi = \xi(s)$ has the expression 
\begin{equation} \label{exsl22}   
\xi (s) = a + \frac{b}{\wp (u_\mathrm{o} +s) - c} \,.   
\end{equation} 
The invariants $g_2$ and $g_3$ of $\wp$ are determined by the coefficients of $f(\xi)$:  
\begin{equation} \label{exsl21} 
g_2 = \beta^2 + \gamma \eta + 3 \Big( \frac{\gamma + \eta}{6} \Big)^2 \,,   
\qquad 
g_3 = - \Big( \frac{\gamma + \eta}{6} \Big) \Big( \beta^2 + \gamma \eta - \Big( \frac{\gamma + \eta}{6} \Big)^2 \Big) \,. 
\end{equation} 
A characterization of the constant $u_\mathrm{o}$ is obtained from the knowledge of the values of $u$ that furnish the roots of $f = f(\xi(u))$ and, with reference to the function $u (s) = u_\mathrm{o} + s$, of the interval, parallel to the real axis in the complex plane, on which $u$ varies for $s$ in $(0, \ell)$. 
Let $\omega_1$ and $\omega_3$ be the real and imaginary half-periods of $\wp$, and let $\omega_2 = \omega_1 + \omega_3$. 
Equations \eqref{exsl16} and \eqref{exsl19} show that the roots of $f(\xi(u))$ correspond to $u=0$, where $\wp$ is infinite, and to $u$ equal to $\omega_1$, $\omega_2$, and $\omega_3$, where $\wp$ vanishes; moreover, equation \eqref{exsl16} implies that $\xi$ is infinite for $u = v = \wp^{-1}c$. 
In view of Equation \eqref{exsl8}, the boundary condition \eqref{exsl1}$_2$ prescribes that 
\begin{equation} \label{exsl23}
\gamma = \eta \xi^2(\ell) + 2 \beta \xi(\ell) \,, 
\end{equation} 
and shows that $\xi(\ell)$ must be equal to one of the roots of $\hat{f}(\xi)$.  
The interval in which the constant $\gamma$ can take its values is determined by the condition that the roots of $\hat{f}(\xi)$ be real and distinct, \textit{i.e.}, $\Delta = \beta^2 + \gamma \eta>0$, which implies
\begin{equation} \label{exsl24} \nonumber  
\gamma > - \frac{\beta^2}{\eta} = \gamma_1 \,,
\end{equation} 
and by the condition \eqref{exsl23} which, by \eqref{exsl15}$_1$, shows that  
\begin{equation} \label{exsl25} \nonumber  
\gamma  < \eta + 2 \beta = \gamma_2 \,,  
\end{equation} 
(the possibility that $\gamma$ be equal to $\gamma_2$ is excluded by the requirement that the polynomial $f(\xi)$ have distinct roots). 
From \eqref{exsl14} and \eqref{exsl18}$_1$ we see that $\xi_a$, $\xi_b$, and $b$ are monotone functions of $\gamma$; those same equations imply that, for $\gamma \to \gamma_1$,  
\begin{equation} \label{exsl26} \nonumber   
\xi_a \to -\frac{\beta}{\eta} \,, \qquad \xi_b \to -\frac{\beta}{\eta} \,, \qquad b \to \frac{(\beta+\eta)^2}{2 \eta} > 0 \,, 
\end{equation} 
and, for $\gamma \to \gamma_2$, 
\begin{equation} \label{exsl27} \nonumber   
\xi_a \to -\Big( 1+ \frac{2\beta}{\eta}\Big)  \,, \qquad \xi_b \to 1 \,, \qquad b \to 0 \,.   
\end{equation} 
Thus, $\xi_a$ can have values less than $-1$, $\xi_b$ is less than 1, and this means that $\xi=1$ is the greatest of the roots of $f$, and $\xi=-1$ or $\xi_a$ is the smallest; moreover, $b$ is always positive.
The derivative with respect to $\varkappa$ of $\xi = a + b / (\varkappa-c)$ is  
\begin{equation} \label{exsl28} 
\frac{d \xi}{d \varkappa} = \frac{-b}{(\varkappa - c )^2} \,.    
\end{equation} 
Let $\mathscr{P}$ be the perimeter of the rectangle of half-periods of $\wp$; since, when $u$ varies on $\mathscr{P}$ starting from $0$ and moving counterclockwise, $\wp (u)$ varies monotonically from $+\infty$ to $-\infty$, from \eqref{exsl28} we see that $\xi$ is a monotone function of $u$ (for $u$ on $\mathscr{P}$). 
To determine the position on $\mathscr{P}$ of the point $v$ where $\xi$ becomes infinite, we calculate $\wp^\prime v$ from the differential equation satisfied by  $\wp$, making use of \eqref{exsl18}$_2$ and \eqref{exsl21}:   
\begin{equation} \label{exsl29} 
\wp^\prime v = \pm \sqrt{4 \wp^3 v - g_2 \wp v - g_3} = \pm \sqrt{\eta b^2} \,.   
\end{equation} 
The fact that $\wp^\prime v$ is real implies that $v$ can be on the segment $(0, \omega_1)$, where  $\wp^\prime u<0$, or on the segment $(\omega_3, \omega_2)$, where $\wp^\prime u>0$. 
As $\xi(u)|_{u=0}=1$ is the greatest root of $f(\xi)$, no root is given by $u$ varying between $0$ and $v$.  
We conclude that $v$ is in $(0, \omega_1)$, and that $\xi(\omega_1)$ is the smallest of the roots, equal to the smaller among $\xi=-1$ and $\xi_a$.
Consequently, the radical in \eqref{exsl29} has the minus sign and, since the values that $\xi$ assumes on $(0, \omega_1)$ do not belong to the admissible interval $[-1,1]$, in the present problem $u (s) = u_\mathrm{o} + s$ must be on the line through $\omega_2$ and $\omega_3$.    
As already observed, equation \eqref{exsl23} shows that $\xi(\ell)$ must be equal to one of the roots of $\hat{f}(\xi)$; if $\xi(\bar{\omega})$ is the value of $\xi$ at the free end of the rod, then  
\begin{equation} \label{exsl30} \nonumber  
\xi(s)_{s=\ell} = \xi(u)_{u=u_\mathrm{o}+\ell} = \xi(u)_{u=\bar{\omega}} \,,  
\end{equation} 
that is,  
\begin{equation} \label{exsl31} \nonumber  
u_\mathrm{o} = \bar{\omega} - \ell \,.
\end{equation} 
From the above discussion we conclude that: 
\textit{i}) if $\xi_a<-1$, then $\xi(\ell)$ is equal to $\xi(\omega_3) = \xi_b$ and $u_\mathrm{o} = \omega_3 - \ell$; 
\textit{ii}) if $\xi_a>-1$, then $\xi(\ell)$ can be equal to $\xi(\omega_3) = \xi_b$ or $\xi(\omega_2)=\xi_a$; it is $u_\mathrm{o} = \omega_3 - \ell$ if $\xi$ if an increasing function of $s$, and $u_\mathrm{o} = \omega_2 - \ell$ if $\xi$ is a decreasing function of $s$.  

To find the value of the integration constant $\gamma$, we note that, as shown by equations \eqref{exsl21}, the invariants $g_2$ and $g_3$ of $\wp$, and hence $u_\mathrm{o}$, depend on that constant; to stress this dependence, we can write $\xi = \xi(s; \gamma)$. 
The value or the values of $\gamma \in ]\gamma_1, \gamma_2[$ that, for assigned $\beta$ and $\eta$, correspond to a solution of the considered equilibrium problem, must satisfy the condition $\xi(s; \gamma)|_{s=0} = \cos \vartheta_\mathrm{o}$.  

The coordinate $y$ of the points of $\mathscr{C}$ follows from the integration of $\xi = y^\prime$. Equation \eqref{exsl22} can be written as 
\begin{equation} \label{exsl33} 
\xi (s) = a + \frac{b}{\wp^\prime v} \, \frac{\wp^\prime v}{\wp (u_\mathrm{o} +s) - \wp v} \,,   
\end{equation} 
where $\wp^\prime v = - b \sqrt{\eta}$; 
using the following addition formula involving the Weierstrass functions $\wp$ and $\zeta$,  
\begin{equation} \label{exsl34} 
\frac{\wp^\prime u}{\wp u - \wp v} = \zeta (u+v) + \zeta (u-v) - 2 \zeta u \,,
\end{equation} 
equation \eqref{exsl33} becomes 
\begin{equation} \label{exsl35} \nonumber   
\xi (s) = a - \frac{1}{\sqrt{\eta}} (\zeta (u_\mathrm{o} + s +v) - \zeta (u_\mathrm{o} + s -v) - 2 \zeta v) \,,    
\end{equation} 
and since $\zeta$ is the logarithmic derivative of the Weierstrass function $\sigma$, taking in account that $y(s)|_{s=0} = 0$, we have  
\begin{equation} \label{exsl36}   
y(s) = \left[ \Big (a - \frac{2 \zeta v}{\sqrt{\eta}} \Big) \varsigma + \frac{1}{\sqrt{\eta}} \ln \frac{\sigma (u_\mathrm{o} + \varsigma + v)}{\sigma (u_\mathrm{o} + \varsigma - v)} \right]_{\varsigma=0}^{\varsigma =s} \,. 
\end{equation} 
To determine the coordinate $z$ of the points of $\mathscr{C}$, we make use of equation \eqref{exsl9a} that, by means of \eqref{exsl8}, is written as    
\begin{equation} \label{exsl37} \nonumber  
y^{\prime \prime} = - \kappa_x z^\prime = \pm z^\prime \sqrt{ \gamma - \eta \xi^2 - 2 \beta \xi} \,, 
\end{equation} 
and implies that    
\begin{equation} \label{exsl38} 
z(\xi) = \pm \int \limits \frac{d \xi}{\sqrt{\gamma - \eta \xi^2 - 2 \beta \xi}} \,. 
\end{equation} 
The term under the radical sign can be put in the form 
\begin{equation} \label{exsl39} \nonumber  
\gamma - \eta \xi^2 - 2 \beta \xi = \frac{\beta^2 + \gamma \eta}{\eta} \Big( 1 - \Big( \frac{\eta \xi + \beta}{\sqrt{\beta^2 + \gamma \eta}} \Big)^2 \Big) \,,    
\end{equation} 
and equation \eqref{exsl38} becomes   
\begin{equation} \label{exsl40} \nonumber  
\begin{split} 
z(\xi) &= \pm \int \limits \frac{d \xi}{\sqrt{\gamma - \eta \xi^2 - 2 \beta \xi}} 
\\
&= \pm \frac{1}{\sqrt{\eta}} \int \Big( 1 - \Big( \frac{\eta \xi + \beta}{\sqrt{\beta^2 + \gamma \eta}} \Big)^2 \Big)^{-\frac{1}{2}} \, d \Big( \frac{\eta \xi + \beta}{\sqrt{\beta^2 + \gamma \eta}} \Big) \,.
\end{split}     
\end{equation} 
Thus, as $z(s)|_{s=0}=0$, we have    
\begin{equation} \label{exsl41} 
z(s) = \pm \frac{1}{\sqrt{\eta}} \left[ \sin^{-1} \Big( \frac{\eta \xi(\varsigma) + \beta}{\sqrt{\beta^2 + \gamma \eta}} \Big)  \right]_{\varsigma=0}^{\varsigma =s} \,,
\end{equation} 
where $\xi(s)$ is given by \eqref{exsl22}.

\remark
If the rod is subject only to the action of the uniform field $\mathbbm{h}_\mathrm{a}$, the equilibrium equation and the quantities that characterize the solution can be obtained by putting $\beta =0$ into the equations of the case with non-null force.  
In particular, the polynomials $f(\xi)$ and $\hat{f}(\xi)$ have the expressions  
\begin{equation} \label{slch1} \nonumber 
f(\xi) = \eta \xi^4 - (\gamma + \eta) \xi^2 + \gamma \,, \qquad \hat{f} (\xi) = - \eta \xi^2 + \gamma \,,   
\end{equation} 
and the roots of $\hat{f}(\xi)$ are 
\begin{equation} \label{slch2} \nonumber 
\xi_a = - \sqrt{\gamma/\eta} \,, \qquad \quad \xi_b = \sqrt{\gamma/\eta} \,. 
\end{equation} 
The boundary condition \eqref{exsl1}$_2$ requires that   
\begin{equation} \label{slch3} \nonumber 
\gamma = \eta \xi^2(\ell)  \,, 
\end{equation} 
and shows that $\gamma$ is positive and less than $\eta$ (the four roots of $f$ are assumed to be distinct), so that $-1<\xi_a<\xi_b<1$.  
The function $\xi$, solution of the differential equation \eqref{exsl12}, has the form  \eqref{exsl22} in which, when $a=1$, the constants $b$ and $c$ have the expressions that can be obtained by putting $\beta=0$ in equations \eqref{exsl18}; analogously, the invariants of $\wp$ are those given by equation \eqref{exsl21} for $\beta = 0$. 
The constants $\gamma$ and $u_\mathrm{o}$ are determined as for the case with non-null force.
The coordinate $y$ of the points of the deformed rod axis is given by equation \eqref{exsl36}, the coordinate $z$ has the expression obtained by setting $\beta=0$ in equation \eqref{exsl41}.  

\remark
If the only action on the rod is the terminal force $\boldsymbol{F}$, the polynomial $f(\xi)$ is of the third degree and the differential equation for $\xi^\prime$ is
\begin{equation} \label{slch4}
(\xi^\prime)^2 = f(\xi) = 2 \beta \xi^3 - \gamma \xi^2 - 2 \beta \xi + \gamma \,.  
\end{equation} 
The roots of $f$ are $\xi_1 = 1$, $\xi_2 = \gamma/(2 \beta)$, and $\xi_3 = -1$. 
The expression of $\kappa_x^2 = (\boldsymbol{d}_3 \times \boldsymbol{d}_3^{\, \prime})^2$ is 
\begin{equation} \label{slch5}
\kappa_x^2 = \gamma - 2\beta \xi(\ell) \,,
\end{equation} 
so that the boundary condition at $s= \ell$ is  
\begin{equation} \label{slch6}
\gamma = 2\beta \xi(\ell) \,.
\end{equation} 
This equation implies that $\gamma_1 = - 2\beta$ and $\gamma_2 = 2\beta$ and, thus, that $\xi_1 > \xi_2 > \xi_3$ (the roots are assumed to be distinct). 
The form of equation \eqref{slch4} that is soluble by means of elliptic functions is obtained by means of the substitution (cf. \cite{LB1930}, Sect.124)
\begin{equation} \label{slch7}
\xi(u) = \frac{\gamma}{6 \beta} + \frac{2}{\beta} \varkappa (u) \,;  
\end{equation} 
assuming that $u = u_\mathrm{o} + s$, introduction of \eqref{slch7} into \eqref{slch4} furnishes  
\begin{equation} \label{slch8} \nonumber 
\Big( \frac{d \xi}{ds} \Big)^2 = \frac{4}{\beta^2} \Big( \frac{d \varkappa}{du} \frac{d u}{ds}  \Big)^2 = \frac{4}{\beta^2} \Big( 4 \varkappa^3 - \Big( \beta^2 + \frac{\gamma^2}{12} \Big) \varkappa - \frac{\gamma^3}{216} + \frac{\beta^2 \gamma}{6} \Big) \,. 
\end{equation} 
This equation shows that $\varkappa$ is the function $\wp$ of Weierstrass, 
\begin{equation} \label{slch9} \nonumber  
\varkappa (u) = \wp (u; g_2, g_3) =  \wp u \,,  
\end{equation} 
whose invariants are 
\begin{equation} \label{slch10} \nonumber 
g_2 = \beta^2 + \frac{\gamma^2}{12} \,, \qquad \quad g_3 = \frac{\gamma^3}{216} - \frac{\beta^2 \gamma}{6} \,;  
\end{equation} 
hence, 
\begin{equation} \label{slch11}
\xi(s) = \frac{\gamma}{6 \beta} + \frac{2}{\beta} \wp (u_\mathrm{o} + s) \,.   
\end{equation} 
As to the constant $u_\mathrm{o}$, we observe that $\lim_{\xi \to \pm \infty} f(\xi) = \pm \infty$, and this means that, in the interval $[-1,1]$, $f$ has positive values between $\xi_3$ and $\xi_2$; since the boundary condition \eqref{slch6} requires that $\xi(\ell)$ be equal to $\xi_2 = \xi (\omega_2)$, we conclude that $u_\mathrm{o} = \omega_2 - \ell$.  
The constant $\gamma$ must satisfy the condition $\xi(s; \gamma)|_{s=0} = \cos \vartheta_\mathrm{o}$ (that we have written emphasizing the dependence of $\xi$ on $\gamma$). 
Integration of $y^\prime = \xi$ with the condition $y(s)|_{s=0} = 0$ yields
\begin{equation} \label{slch12} \nonumber 
y(s) = \left[  \frac{\gamma}{6 \beta} \varsigma - \frac{2}{\beta} \zeta (u_\mathrm{o} + s) \right]_{\varsigma=0}^{\varsigma=s} \,; 
\end{equation} 
From equations \eqref{exsl9a} and \eqref{slch5} it follows that 
\begin{equation} \label{slch13} \nonumber  
z(\xi) = \pm \int \limits \frac{d \xi}{\sqrt{\gamma - 2 \beta \xi}} \,, 
\end{equation} 
whose integration with the condition $z(s)|_{s=0} = 0$ furnishes   
\begin{equation} \label{slch14} \nonumber 
z(s) = \pm \left[ \frac{\sqrt{\gamma-2 \beta \xi(\varsigma)}}{-\beta}  \right]_{\varsigma=0}^{\varsigma=s} \,, 
\end{equation} 
where $\xi(s)$ is given by the equation \eqref{slch11}.

\subsection{Solution for $\mathbbm{h}_\mathrm{a}$ orthogonal to $\boldsymbol{F}$} 

This case can be studied following the same steps as those of the case in which $\mathbbm{h}_\mathrm{a}$ and $\boldsymbol{F}$ are parallel.   
To simplify the writing, we put $\overline{\gamma} = \eta - \gamma$, so that equation \eqref{exsl8} becomes  
\begin{equation} \label{exhpf1} 
\kappa_x^2 = \eta \xi^2 - 2 \beta \xi - \overline{\gamma} \,,   
\end{equation}
and the polynomial $f (\xi)$ and $\hat{f} (\xi)$ have the expressions  
\begin{equation} \label{exhpf2} \nonumber  
f(\xi) = - \eta \xi^4 + 2 \beta \xi^3 + (\overline{\gamma} + \eta) \xi^2 - 2 \beta \xi - \overline{\gamma} \,, \qquad \hat{f} (\xi) = \eta \xi^2 - 2 \beta \xi - \overline{\gamma} \,.  
\end{equation}
The roots $\xi_a$ and $\xi_b$ of $\hat{f} (\xi)$, with $\xi_a < \xi_b$, are 
\begin{equation} \label{exhpf3}
\left. \begin{matrix}
\xi_a \\ \xi_b  
\end{matrix} \right\} 
= \frac{\beta \mp \sqrt{\beta^2 + \overline{\gamma} \eta}}{\eta} = \frac{\beta}{\eta} \Big( 1 \mp \sqrt{1 + \frac{\overline{\gamma} \eta}{\beta^2}} \Big)  \,,  
\end{equation} 
and, since the coefficient of $\xi^2$ in $\hat{f} (\xi)$ is positive, $f(\xi) =\hat{f} (\xi) (1-\xi^2)$ has nonnegative values for $\xi$ not belonging to the interior of the interval of the roots of $\hat{f} (\xi)$; the admissible values of $\xi$ must satisfy the relations 
\begin{equation}  \label{exhpf4} 
\xi \in [ -1, 1 ] \,, \qquad \xi \notin \, ] \xi_a, \xi_b [ \,. 
\end{equation} 
The solution of the equation \eqref{exsl12} again has the form 
\begin{equation} \label{exhpf5}   
\xi (s) = a + \frac{b}{\wp (u_\mathrm{o} +s) - c} \,,    
\end{equation} 
where, for $a=1$, now it is 
\begin{equation} \label{exhpf6} 
b = \beta + \frac{\overline{\gamma} - \eta}{2} \,, \qquad \quad c = \frac{\beta - \eta}{2} + \frac{\overline{\gamma} + \eta}{12} \,,   
\end{equation} 
and the invariants $g_2$ and $g_3$ of $\wp$ are:  
\begin{equation} \label{exhpf7} \nonumber  
g_2 = \beta^2 + \overline{\gamma} \eta + 3 \Big( \frac{\overline{\gamma} + \eta}{6} \Big)^2 \,,   
\quad 
g_3 = \Big( \frac{\overline{\gamma} + \eta}{6} \Big) \Big( \beta^2 + \overline{\gamma} \eta - \Big( \frac{\overline{\gamma} + \eta}{6} \Big)^2 \Big) \,. 
\end{equation} 
Equations \eqref{exhpf3} and \eqref{exhpf6}$_1$ show that $\xi_a$, $\xi_b$, and $b$ are monotone functions of $\overline{\gamma}$. 
The ends of the interval of the possible values of $\overline{\gamma}$, determined as previously done, are:  
\begin{equation} \label{exhpf8} \nonumber 
\overline{\gamma}_1 = - \frac{\beta^2}{\eta} \,, \qquad \quad \overline{\gamma}_2 = \eta + 2 \beta \,; 
\end{equation} 
we find that, for $\overline{\gamma} \to \overline{\gamma}_1$, 
\begin{equation} \label{exhpf9} \nonumber 
\xi_a \to \frac{\beta}{\eta} \,, \qquad \xi_b \to \frac{\beta}{\eta} \,, \qquad b \to - \frac{(\beta-\eta)^2}{2 \eta} \,, 
\end{equation} 
and, for $\overline{\gamma} \to \overline{\gamma}_2$, 
\begin{equation} \label{exhpf10} \nonumber 
\xi_a \to -1 \,, \qquad \xi_b \to \xi_b = 1 + \frac{2 \beta}{\eta} \,, \qquad b \to 2 \beta \,. 
\end{equation} 
We see that $\xi_a$ is always greater than $-1$ and $\xi_b$ is positive, so that $\xi=-1$ is the smallest of the roots of $f$, and $\xi=1$ or $\xi_b$ is the greatest; moreover, $b$ can be positive or negative and the expression \eqref{exsl28} of the derivative of $\xi$ with respect to $\varkappa = \wp u$, shows that $\xi$ can be an increasing or a decreasing function of $u$, when $u$ moves  along the perimeter $\mathscr{P}$ of the rectangle of the half-periods. 
The derivative of $\wp$ at the point $v = \wp^{-1}c$, where $\xi$ becomes infinite, is 
\begin{equation} \label{exhpf11} \nonumber   
\wp^\prime v = \pm i \sqrt{\eta b^2} \,;    
\end{equation} 
thus, in view of the properties of the function $\wp$, the point $v$ is on the segment from $\omega_3$ to $0$ if $\wp^\prime v$ is imaginary negative, and is on the segment from $\omega_1$ to $\omega_2$ if $\wp^\prime v$ is imaginary positive. 
The boundary condition for $s=\ell$ requires that 
\begin{equation} \label{exhpf12} \nonumber  
\overline{\gamma} = \eta \xi^2(\ell) - 2 \beta \xi (\ell)  \,,   
\end{equation}
with $\xi(\ell)$ equal to one of the roots of $\hat{f}(\xi)$. 
Taking into account that the root $\xi = 1$ occurs at $u=0$ and that no root is given by $u$ varying between $v$ and the point $\bar{u}$ for which $\xi(u)|_{u=\bar{u}} = -1$, we have: 
\textit{i}) if $v$ is on the segment $(0, \omega_3)$, 1 is the greatest root, and $\xi_b = \xi (\omega_1)$, $\xi_a = \xi(\omega_2)$, $-1 = \xi(\omega_3)$; if $\xi(s)|_{s=0} = \cos \vartheta_\mathrm{o} > \xi_b$, $u = u_\mathrm{o} + s$ varies on the segment $(0, \omega_1)$ and $u_\mathrm{o} = \omega_1 - \ell$; if $\xi(s)|_{s=0} = \cos \vartheta_\mathrm{o}< \xi_a$, $u = u_\mathrm{o} + s$ varies on the segment $(\omega_3, \omega_2)$ and $u_\mathrm{o} = \omega_2 - \ell$; 
\textit{ii}) if $v$ is on the segment $(\omega_1, \omega_2)$, taking into account equation \eqref{exhpf4}$_2$ and the fact that $\xi(s)|_{s=\ell}$ must be equal to one of the roots of $\hat{f}(\xi)$, we have $-1=\xi(\omega_2)$, $\xi_a = \xi(\omega_3)$, $\xi_b = \xi(\omega_1) > 1$, and $u_\mathrm{o} = \omega_3 - \ell$. 
The value of $\wp^\prime v$ is  
\begin{equation} \label{exhpf13bis} \nonumber 
\wp^\prime v = 
\left\{ \begin{matrix}
- i \sqrt{\eta b^2} \,, \quad \mathrm{if} \,\,\,\, c = \wp v < \wp \omega_3 \,, 
\\ 
+ i \sqrt{\eta b^2} \,, \quad \mathrm{if} \,\,\,\, c = \wp v > \wp \omega_2 \,.   
\end{matrix} \right.     
\end{equation} 
As in the cases previously examined, we can write $\xi = \xi(s; \bar{\gamma})$ and the values of the constant $\overline{\gamma} \in ] \gamma_1, \gamma_2 [$ that correspond to solutions of the equilibrium problem are obtained from the condition $\xi(s; \bar{\gamma})|_{s=0} = \cos \vartheta_\mathrm{o}$. 
By means of the formula \eqref{exsl34}, equation \eqref{exhpf5} can be put in the form 
\begin{equation} \label{exhpf14} \nonumber  
\xi (s) = a - \frac{b}{\wp^\prime v} (\zeta (u_\mathrm{o} + s +v) - \zeta (u_\mathrm{o} + s -v) - 2 \zeta v) \,,    
\end{equation} 
which, by integration with the condition $y(s)|_{s=0} = 0$, furnishes  
\begin{equation} \label{exhpf15} \nonumber    
y(s) = \left[ \Big (a + \frac{2 b \zeta v}{\wp^\prime v} \Big) \varsigma - \frac{b}{\wp^\prime v} \ln \frac{\sigma (u_\mathrm{o} + \varsigma + v)}{\sigma (u_\mathrm{o} + \varsigma - v)} \right]_{\varsigma=0}^{\varsigma =s} \,. 
\end{equation} 
Equations \eqref{exsl9a} and \eqref{exhpf1} yield   
\begin{equation} \label{exhpf16}  
z(\xi) = \pm \int \limits \frac{d \xi}{\sqrt{\eta \xi^2 - 2 \beta \xi - \overline{\gamma}}} \,;  
\end{equation} 
the term under the radical sign can be written as 
\begin{equation} \label{exhpf17} \nonumber  
\eta \xi^2 - 2 \beta \xi - \overline{\gamma} = \frac{\beta^2 + \overline{\gamma} \eta}{\eta} \Big( \Big( \frac{\eta \xi - \beta}{\sqrt{\beta^2 + \overline{\gamma} \eta}} \Big)^2 - 1 \Big) \,,   
\end{equation} 
and \eqref{exhpf16} becomes  
\begin{equation} \label{exhpf18} \nonumber  
\begin{split} 
z(\xi) &= \pm  \int \limits \frac{d \xi}{\sqrt{\eta \xi^2 - 2 \beta \xi - \overline{\gamma}}} 
\\
&= \pm \frac{1}{\sqrt{\eta}} \int \Big( \Big( \frac{\eta \xi - \beta}{\sqrt{\beta^2 + \overline{\gamma} \eta}} \Big)^2 -1 \Big)^{-\frac{1}{2}} \, d \Big( \frac{\eta \xi - \beta}{\sqrt{\beta^2 + \overline{\gamma} \eta}} \Big)  \,.    
\end{split} 
\end{equation} 
From this equation, taking into account that $z(s)|_{s=0}=0$, we obtain      
\begin{equation} \label{exhpf19} \nonumber  
z(s) = \pm \frac{1}{\sqrt{\eta}} \left[ \cosh^{-1} \Big( \frac{\eta \xi(\varsigma) - \beta}{\sqrt{\beta^2 + \gamma \eta}} \Big)  \right]_{\varsigma=0}^{\varsigma =s} \,,
\end{equation} 
where $\xi(s)$ is given by \eqref{exhpf5}.

\section{Examples}

Two examples of application of the exact solutions deduced in the previous Section are presented: the first one considers a problem in which the force and the magnetic field are, respectively, orthogonal and parallel to the direction of the undeformed rod axis; the second one refers to a problem in which the directions of the force applied at the free end and the magnetic field are parallel.

\subsection{Example in which $\mathbbm{h}_\mathrm{a}$ and $\boldsymbol{F}$ are orthogonal}

\begin{figure}[h!]   
\begin{center}
\begin{tikzpicture}[>=angle 90,scale=1.0]]
\tikzset{ground2/.style={fill,pattern=north east lines,draw=none, minimum width=0.05cm,minimum height=0.05cm}}
\tikzset{ground/.style={fill,pattern=north east lines,draw=none, minimum width=0.75cm, minimum height=0.1cm}}
\tikzset{groundv/.style={fill,pattern=north east lines,draw=none, minimum width=0.05cm,minimum height=0.75cm}}
\tikzset{groundv2/.style={fill,pattern=north east lines,draw=none, minimum width=0.05cm,minimum height=0.04cm}}
\pgfmathsetmacro{\lun}{6}
\coordinate (A) at (0,0);
\coordinate (B) at ($(A)+(\lun,0)$);
%
%
\draw[black] (A) -- (B); 
%
%
\node (h1) at (A) [groundv,anchor=east]{};
\draw (h1.north east) -- (h1.south east);
%
%
\coordinate (O) at (0,0);
\draw [black,->] (O) -- ++(-90:1.2) node[left] {$y$};
\draw [black,->] (O)  -- ++(0:1.2) node[below] {$z$};
\draw[black,->] ($(A)+(-90:0.5)$) arc (-90:0:0.5); 
\node[black] at ($(A)+(0.6,-0.5)$) {$\vartheta_\mathrm{o}$};
%
%
\node [black] at ($(A)+(0.6,0.3)$) {$s=0$};
\node [black] at ($(B)+(-0.6,0.3)$) {$s=\ell$};
%
%
\draw [black, thick, ->] ($($(A)!0.50!(B)$)+(-0.75,0.6)$) -- ++(0:1.5);
\draw[black] ($($(A)!0.50!(B)$)+(0.4,1.0)$) node[left] {$\mathbbm{h}_\mathrm{a}$};
%
%
\draw [black, thick, <-] ($(B)+(0.02,0.02)$)--node[right] {$\boldsymbol{F}$} ++(90:1.0);
\end{tikzpicture}
\end{center}
\vspace{-5pt}
\caption{Cantilever subject to a force $\boldsymbol{F}$ orthogonal to the direction of the undeformed rod axis, and immersed in a magnetic field parallel to that direction.}  
\label{Fi:figura_H_F_ort}
\end{figure}
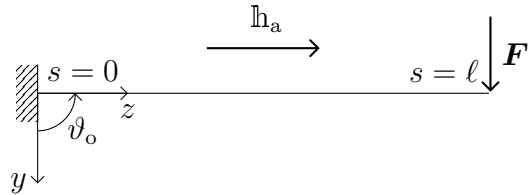

We consider a cantilever that has length $\ell=80$ in dimensionless units, is subject to a terminal force orthogonal to the direction of the undeformed rod axis, and is immersed in a magnetic field parallel to that direction, as shown in the Figure \ref{Fi:figura_H_F_ort}. 
\smallskip 

\noindent  
In this arrangement, the force tends to bend the rod, while the magnetic field produces a distribution of couples that, for the assumed orientation of the magnetic particles, tend to keep the longitudinal fibers of the rod parallel to the field. 
\begin{figure}[h!]
\centering 
\fbox{\includegraphics[scale=0.65]{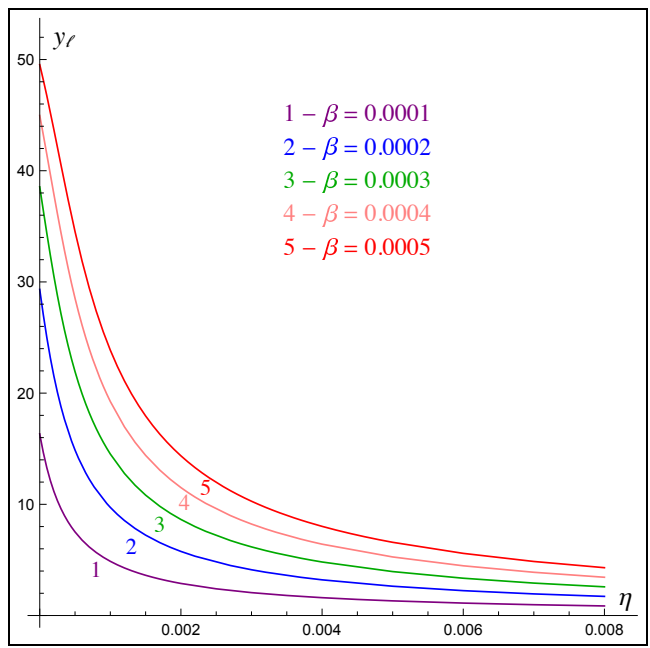}}
\caption{Transverse component of the displacement of the free end as a function of $\eta$ for  some values of $\beta$ in the range from 0.0001 to 0.0005.}  
\label{Fi:F_H_ortogonali} 
\end{figure}

We assume that, initially, the rod is bent by the terminal force in absence of the magnetic field; then, the field is applied in successive steps (avoiding dynamical effects) and produces a progressive reduction of the inflection. 

In the Figure \ref{Fi:F_H_ortogonali}, we have plotted, for increasing values of $\eta$, the coordinate $y_\ell$ of the free end of the rod, which is equal to the transverse component of the displacement at $s = \ell$. 
We considered five values of the force, corresponding to the values of $\beta$ written in the Figure. 
The diagrams show that the action of the magnetic field reduces the displacement to a small amount of its initial value. 
Thus, the problem treated in this example can be viewed as a model for a remotely controlled deformation of a rod. 

\subsection{Example in which $\mathbbm{h}_\mathrm{a}$ and $\boldsymbol{F}$ are parallel}

\begin{figure}[h!]
\centering 
\includegraphics[scale=0.65]{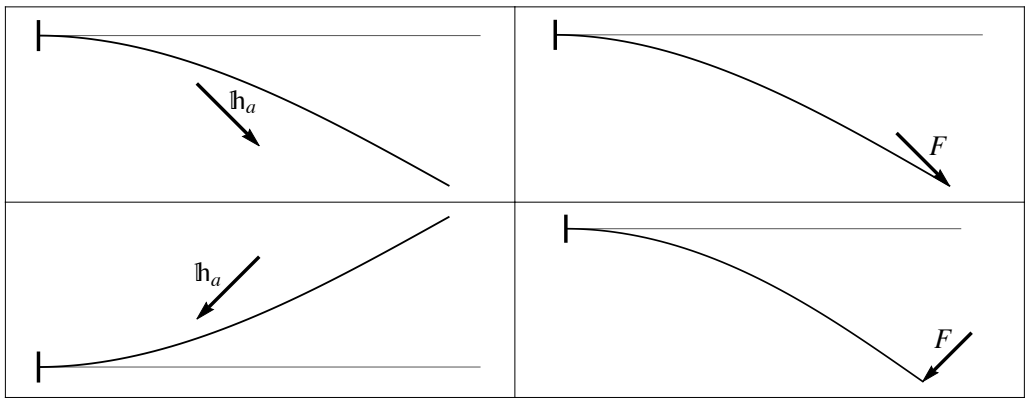}
\caption{Examples of deformations caused by a magnetic field and a terminal force acting separately on a cantilever.}  
\label{Fi:figura_HF} 
\end{figure}

In the present example we refer to a cantilever subject to a magnetic field $\mathbbm{h}_\mathrm{a}$ and a terminal force $\boldsymbol{F}$ having the same direction. 
We note preliminarily that, as the orientation of the magnetic particles is assumed to be parallel to the longitudinal fibers of the rod, the couples exerted by a field $\mathbbm{h}_\mathrm{a}$ tend to make those fibers parallel to the direction of the magnetic field;   
as shown by Figure \ref{Fi:figura_HF}, a magnetic field and a force, having the same direction and acting separately, bend a cantilever on sides of the undeformed rod axis that can be coinciding or opposite.  

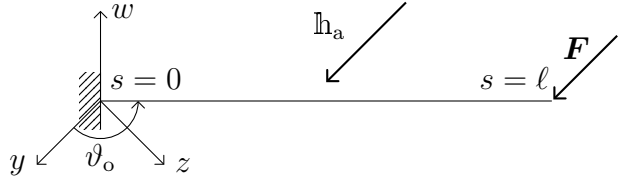
\begin{figure}[h!]   
\begin{center}
\begin{tikzpicture}[>=angle 90,scale=1.0]]
\tikzset{ground2/.style={fill,pattern=north east lines,draw=none, minimum width=0.05cm,minimum height=0.05cm}}
\tikzset{ground/.style={fill,pattern=north east lines,draw=none, minimum width=0.75cm, minimum height=0.1cm}}
\tikzset{groundv/.style={fill,pattern=north east lines,draw=none, minimum width=0.05cm,minimum height=0.75cm}}
\tikzset{groundv2/.style={fill,pattern=north east lines,draw=none, minimum width=0.05cm,minimum height=0.04cm}}
\pgfmathsetmacro{\lun}{6}
\coordinate (A) at (0,0);
\coordinate (B) at ($(A)+(\lun,0)$);
%
%
\draw[black] (A) -- (B); 
%
%
\node (h1) at (A) [groundv,anchor=east]{};
\draw (h1.north east) -- (h1.south east);
%
%
\coordinate (O) at (0,0);
\draw [black,->] (O) -- ++(-135:1.2) node[left] {$y$};
\draw [black,->] (O)  -- ++(-45:1.2) node[right] {$z$};
\draw[black,->] ($(A)+(-135:0.5)$) arc (-135:0:0.5); 
\node[black] at ($(A)+(0.0,-0.75)$) {$\vartheta_\mathrm{o}$};
\draw [black,->] (O)  -- ++(90:1.2) node[right] {$w$};
%
%
\node [black] at ($(A)+(0.6,0.3)$) {$s=0$};
\node [black] at ($(B)+(-0.5,0.3)$) {$s=\ell$};
%
%
\draw [black, thick, <-] ($($(A)!0.50!(B)$)+(0.0,0.25)$) -- ++(45:1.5);
\draw[black] ($($(A)!0.50!(B)$)+(0.4,1.0)$) node[left] {$\mathbbm{h}_\mathrm{a}$};
%
%
\draw [black, thick, <-] ($(B)+(0.02,0.02)$)--++(45:1.2);
\draw ($(B)+(0.65,0.7)$) node[left] {$\boldsymbol{F}$};
\end{tikzpicture}
\end{center}
\vspace{-5pt}
\caption{Cantilever subject to the field $\mathbbm{h}_\mathrm{a}$ and the force $\boldsymbol{F}$ whose directions form an angle $\vartheta_\mathrm{o} = 3 \pi/4$ with the undeformed rod axis.}  
\label{Fi:figura_H_F_par}
\end{figure}

\noindent
We consider the case in which a uniform field $\mathbbm{h}_\mathrm{a}$ of increasing magnitude acts on a rod that is deformed by the action of a force: we keep the force constant and examine the evolution of the deformation produced by the increasing field, that is assumed to be applied in successive steps avoiding dynamical effects. 
In the example, the common direction of $\boldsymbol{F}$ and $\mathbbm{h}_\mathrm{a}$ forms an angle $\vartheta_\mathrm{o} = 3 \pi/4$ with the undeformed rod axis, as illustrated by the Figure \ref{Fi:figura_H_F_par}.  
\noindent

\begin{figure}[h!]
\centering 
\fbox{\includegraphics[scale=0.75]{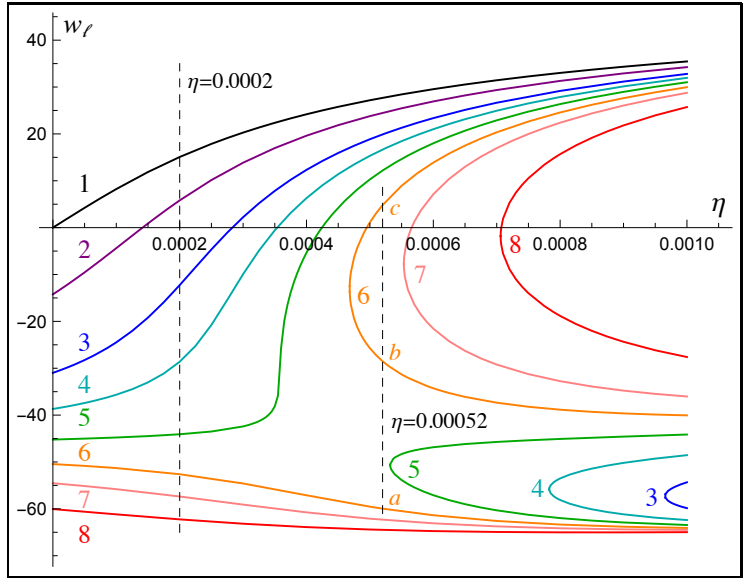}}
\caption{Transverse displacements $w_\ell$ of the free end of a cantilever, whose dimensionless length is 80, that is subject to a constant terminal force and is immersed in a magnetic field of increasing magnitude. 
On each curve the parameter $\beta$ is constant and has the following values: curve 1, $\beta=0$; curve 2, $\beta=0.0001$; curve 3, $\beta=0.0002$; curve 4, $\beta=0.00025$; curve 5, $\beta=0.0003$; curve 6, $\beta=0.00035$; curve 7, $\beta=0.0004$; curve 8, $\beta=0.0005$.}  
\label{Fi:F_fisso_H_variabile} 
\end{figure}

\noindent 
Let $w_\ell$ be the transverse displacement component of the free end of the rod, assumed positive in the upward direction of Figure \ref{Fi:figura_H_F_par}. 
In the Figure \ref{Fi:F_fisso_H_variabile} the diagrams of $w_\ell$ for eight given values of $\beta$ (\textit{i.e.}, of the terminal force) and for increasing values of $\eta$ (\textit{i.e.}, of $| \mathbbm{h}_\mathrm{a} |$) are drawn. 

\noindent 
The diagrams show that the application of an increasing magnetic field can produce either a progressive reduction, followed by an inversion of sign, of the deformation caused by the force, or an increment of that deformation. 
The former effect occurs for lower, and the latter for higher, values of the force. 
The diagrams show also that, as the parameter $\eta$ increases, the equilibrium problems exhibit bifurcation points (\textit{i.e.}, points where a change in the number of solutions occurs) at which couples of new solutions appear; the bifurcation points of the curves 1 and 2 are outside the region represented in the Figure. 
For the considered class of problems, Figure \ref{Fi:F_fisso_H_variabile} may be seen as a bifurcation diagram with $\eta$ as a bifurcation parameter and $\beta$ as an imperfection parameter.   

The plots in the Figure \ref{Fi:F_fisso_H_variabile} show that the behavior of the curves 1-5, corresponding to lower values of the force, is qualitatively different from that of the curves 6-8, corresponding to higher values of the force; the transition from one type of behavior to the other occurs in the region between the curves 5 and 6, that is, for a value of $\beta$ between 0.00030 and 0.00035.  
\begin{figure}[h!]
\centering 
\includegraphics[scale=0.60]{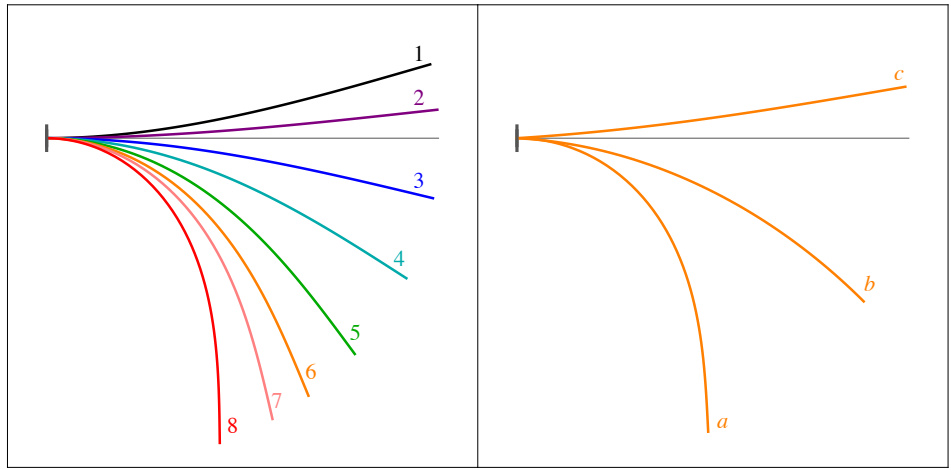}
\caption{Deformed axial curves, at $\eta=0.0002$, for the eight chosen values of $\beta$ (left); deformed axial curves, at $\eta=0.00052$, of the three solutions of the case in which $\beta=0.00035$ (right).}  
\label{Fi:deformate} 
\end{figure}

\noindent 
On the left in Figure \ref{Fi:deformate}, the deformed axial curves of the cantilever at $\eta=0.0002$, for the eight values of $\beta$, are shown. 
On the right in the Figure, with reference to the curve labeled with the number 6, the axial curves in the three equilibrium configurations of the rod for $\eta = 0.00052$ are shown.

\section{Conclusions} 

We have considered magneto-elastic rods whose structure presents a uniform distribution of paramagnetic particles that are firmly embedded in an elastic matrix and are aligned along the longitudinal fibers of the rod.
With reference to rods of that type, viewed as three-dimensional bodies, the expression of the distribution of couples per unit length of the rod axis has been derived. 
The deduction, which is based on an order of magnitude analysis and considers spatial deformations of rods that may be non-prismatic in an undistorted configuration, yields results that are fully consistent with Kirchhoff's theory of rods because they hold at the same order of magnitude.   
Then, by means of a variational procedure, the equilibrium equations of magneto-elastic rods subject to end loads and a uniform magnetic field have been derived. 

The deduced equations have been applied to the study of equilibrium of cantilevers that undergo planar deformation under the action of a terminal force and a magnetic field whose directions are parallel or orthogonal. 
For such classes of problems exact solutions in terms of Weierstrass elliptic functions have been derived, and two examples of applications of these solutions have been presented. 
In the first one a cantilever is subject to a magnetic field parallel to the direction of the undeformed rod axis and a force orthogonal to that direction; the results show that the distribution of couples originated by the magnetic field produces a strong reduction of the deformation caused by the force, and suggest that the considered example constitutes a model of a device for the remote control of the deformation of a rod.  
In the second example the cantilever is acted upon by a force and a magnetic field that have parallel directions forming an angle of $3 \pi/4$ with the undeformed rod axis. 
Curves for constant force and varying magnetic field have been drawn for various values of the force. 
The results show that, as the magnetic field increases, bifurcation points appear; moreover, the qualitative aspect of the bifurcation diagrams is different for lower and higher values of the force. 

The results obtained in the paper suggest that a wide variety of situations can be expected by varying the parameters entering in the equilibrium and stability problems (dimensions of the rod, boundary conditions, magnitude of the force and the magnetic field, inclination of their direction with respect to the undeformed rod axis) and the deduced exact solutions appear to be a useful instrument to investigate these problems.


\begin{thebibliography}{00}

%
\bibitem[Bianchi(1930)]{LB1930}
Bianchi L., 1930.      
Lezioni sulla Teoria delle Funzioni di Variabile Complessa e delle Funzioni Ellittiche, Vol.2. Zanichelli, Bologna.   
%
%
\bibitem[Cebers and Cirulis(2007)]{CC2007}
Cebers A., Cirulis T., 2007.  
Magnetic elastica. 
Phys. Rev. E. 76, 031504. 
%
%
\bibitem[Ciambella et al.(2017)]{CF2017}
Ciambella J., Favata A., Tomassetti G., 2017. 
A nonlinear theory for fiber-reinforced magneto-elastic rods.
Proc. R. Soc. A. 474, 201730703. 
%
%
\bibitem[Clebsch(1862)]{CA1862}
Clebsch A., 1862.  
Theorie der Elasticit\"{a}t Fester K\"{o}rper. B.G. Teubner, Leipzig.
%
%
\bibitem[Clebsch(1883)]{CA1883}
Clebsch A., 1883.  
Th\'eorie de l'Elasticit\'e des Corps Solides, translation of \cite{CA1862} by Saint-Venant and Flamant. 
Dunod, Paris.
%
%
\bibitem[Coleman et al.(1993)]{BC1993}
Coleman B.D., Dill E.H., Lembo M., Lu Z., Tobias I., 1993.      
On the dynamics of rods in the theory of Kirchhoff and Clebsch.    
Arch. Rational Mech. Anal. 121, 339-359.   
%
%
\bibitem[Coleman and Swigon(2000)]{CS2000}
Coleman B.D., Swigon D., 2000.   
Theory of supercoiled elastic rings with self-contact and its application to DNA plasmids.  
J. Elasticity, 60, 173-221. 
%
%
\bibitem[Dill(1992)]{ED1992}
Dill E.H., 1992.  
Kirchhoff's Theory of Rods.  
Arch. Hist. of Exact Sci. 44, 1-23. 
%
%
\bibitem[Dreyfus et al.(2005)]{DB2005}
Dreyfus R., Baudry J., Roper M.L., Fermigier M., Stone H.A., Bibette J., 2005. 
Microscopic artificial swimmers. 
Nature, 437, 862-865. 
%
%
\bibitem[Durastanti et al.(2020)]{DGT2020}
Durastanti R., Giacomelli L., Tomassetti G., 2020. 
Shape programming of a magnetic elastica. 
arXiv: 2003.02696 [math.AP]. 
%
%
\bibitem[Gerbal et al.(2015)]{GW2015}
Gerbal F., Wang Y., Lyonnet F., Bacri J.-C., Hocquet T., Devaud M., 2015. 
A refined theory of magnetoelastic buckling matches experiments with ferromagnetic and superparamagnetic rods. 
PNAS, 112, 7135-7140. 
%
%
\bibitem[Goubault et al.(2003)]{GF2003}
Goubault C., Jop P., Fermigier M., Baudry J., Bertrand E., Bibette J., 2003.  
Flexible magnetic filaments as micromechanical sensors.  
Phys. Rev. Lett. 91, 260802. 
%
%
\bibitem[Hubert and Sch{\"a}fer(1998)]{HS1998}
Hubert A., Sch{\"a}fer R., 1998. 
Magnetic domains: the analysis of magnetic microstructures. 
Springer, New York.
%
%
\bibitem[Kimura et al.(2012)]{KU2012}
Kimura T., Umehara Y., Kimura F., 2012. 
Magnetic field responsive silicone elastomer loaded with short steel wires having orientation distribution. 
Soft Matter, 8, 6206-6209. 
%
%
\bibitem[Kirchhoff(1859)]{KG1859}
Kirchhoff G., 1859. 
\"{U}ber das Gleichgewicht und die Bewegung eines unendlich d\"{u}nnen elastichen Stabes. J. f. reine. angew. Math. (Crelle) 56, 285-313.
%
%
\bibitem[Kirchhoff(1876)]{KG1876}
Kirchhoff G., 1876.  
Vorlesungen \"{u}ber mathematische Physik, Mechanik, Vorl.28, B.G. Teubner, Leipzig.
%
%
\bibitem[Lembo(2003)]{ML2003}
Lembo M., 2003.      
On the stability of elastic annular rods.    
Int. J. Solids Struct. 40, 317-330.   
%
%
\bibitem[Lembo(2016)]{ML2016}
Lembo M., 2016.       
On nonlinear deformations of nonlocal elastic rods.     
Int. J. Solids Struct. 90, 215-227.   
%
%
\bibitem[Lembo(2017)]{ML2017}
Lembo M., 2017.    
Exact solutions for post-buckling deformations of nanorods.     
Acta Mech. 228, 2283-2298.   
%
%
\bibitem[Lembo(2018)]{ML2018}
Lembo M., 2018.       
Exact equilibrium solutions for nonlinear spatial deformations of nanorods with application to buckling under terminal force and couple.    
Int. J. Solids Struct.  135, 274-288.   
%
%
\bibitem[Love(1944)]{AL1944} 
Love A.E.H., 1944.       
A Treatise on the Mathematical Theory of Elasticity, reprint of the fourth edition. 
Dover Publications, New York.
%
%
\bibitem[Moon and Holmes(1979)]{MH1979}
Moon F.C., Holmes P., 1979.    
A magnetoelastic strange attractor.  
J. Sound Vib. 65, 275-296.    
%
%
\bibitem[Moon and Pao(1969)]{MP1969}
Moon F.C., Pao Y.H., 1969.  
Vibration and dynamic instability of a beam-plate in a transverse magnetic field.      
J. Appl. Mech. 36, 92-100. 
%
%
\bibitem[Seidman and Wolfe(1988)]{SW1988}
Seidman T.I. and Wolfe P., 1988. 
Equilibrium states of an elastic conducting rod in a magnetic field. 
Arch. Rational Mech. Anal. 102, 308-329. 
%
%
\bibitem[Tiersten(1990)]{HT1990}
Tiersten H. F., 1990.      
A Development of the Equations of Electromagnetism in Material Continua.    
Springer-Verlag, New York.   
%
%
\bibitem[Tobias et al.(1996)]{TCL1996} 
Tobias I., Coleman B.D., Lembo M., 1996.  
A class of exact dynamical solutions in the elastic model of DNA with implications for the theory of fluctuations in the torsional motions of plasmids.  
J. Chem. Phys. 105 (6), 2517-2526.
%
%
\bibitem[Tobias et al.(1994)]{TCO1994}
Tobias I., Coleman B.D., Olson W.K.., 1994.  
The dependence of DNA tertiary structure on end conditions: Theory and implications for topological transitions. 
J. Chem. Phys. 101, 10990-10996. 
%
%
\bibitem[Vella et al.(2013)]{VP2013} 
Vella D., du Pontavice E., Hall C.L., Goriely A., 2013.  
The magneto-elastica: from self-buckling to self-assembly. 
Proc. R. Soc. A, 470, 20130609. 
%
%
\bibitem[Wallerstein and Peach(1972)]{WP1972}
Wallerstein D.V., Peach M.O., 1972.  
Magnetoelastic buckling of beams and thin plates of magnetically soft material,
J. Appl. Mech., 39, 451-455. 
%
%
\bibitem[Wang et al.(2020)]{WK2020} 
Wang L., Kim Y., Guo C. F., Zhao X., 2020. 
Hard-magnetic elastica. 
J. Mech. Phys. Solids, 142, 104045. 
%
%
\bibitem[Wolfe(1983)]{PW1983}
Wolfe P., 1983.  
Equilibrium states of an elastic conducting wire in a magnetic field: a paradigm of bifurcation theory. 
T. Am. Math. Soc. 278, 377-387. 
%
%
\end{thebibliography}
\end{document}